\documentclass[aps,pre,amsmath,amssymb,lengthcheck,showpacs,superscriptaddress]{revtex4-1}
\usepackage{graphicx}
\usepackage{color}

\begin{document}

\title{Anomalous vibrational properties in the continuum limit of glasses}

\author{Masanari Shimada}
\affiliation{Graduate School of Arts and Sciences, The University of Tokyo, Tokyo 153-8902, Japan}
\author{Hideyuki Mizuno}
\affiliation{Graduate School of Arts and Sciences, The University of Tokyo, Tokyo 153-8902, Japan}
\author{Atsushi Ikeda}
\affiliation{Graduate School of Arts and Sciences, The University of Tokyo, Tokyo 153-8902, Japan}

\date{\today}

\begin{abstract}
The low-temperature thermal properties of glasses are anomalous with respect to those of crystals. 
These thermal anomalies indicate that the low-frequency vibrational properties of glasses differ from those of crystals. 
Recent studies revealed that, in the simplest model of glasses, i.e., the harmonic potential system, phonon modes coexist with soft localized modes in the low-frequency (continuum) limit. 
However, the nature of low-frequency vibrational modes of more realistic models is still controversial.
In the present work, we study the Lennard-Jones (LJ) system using large-scale molecular-dynamics (MD) simulation and establish that the vibrational property of the LJ glass converges to coexistence of the phonon modes and the soft localized modes in the continuum limit as in the case of the harmonic potential system.
Importantly, we find that the low-frequency vibrations are rather sensitive to the numerical scheme of potential truncation which is usually implemented in the MD simulation, and this is the reason why contradictory arguments have been reported by previous works. 
We also discuss the physical origin of this sensitiveness by means of a linear stability analysis.
\end{abstract}

\maketitle

\section{Introduction}
The low-temperature thermal properties of glasses are anomalous with respect to their crystalline counterparts~\cite{phillips1981amorphous}. 
The specific heat of crystals follows $C \propto T^3$ and the thermal conductivity follows $\kappa \propto T^3$ at low temperatures $T$, regardless of their constituent particles~\cite{kittel1996introduction}.
These universal behaviors are explained by theories based on quantized lattice vibrations or phonons; for example, the Debye theory can predict the vibrational density of states (vDOS) $g(\omega)\propto \omega^2$ ($\omega$ is frequency) and then the specific heat $C\propto T^3$. 
In contrast, glasses exhibit different universal behaviors~\cite{phillips1981amorphous}.
The specific heat of glasses is larger than corresponding crystalline value around $T \sim 10$~[K]~\cite{zeller1971thermal}. 
This is a direct reflection of the excess vibrational modes over the prediction of the Debye theory around $\omega_{\mathrm{BP}} \sim 1$~[THz], the so-called boson peak (BP)~\cite{buchenau1984neutron}.  
At lower temperatures, $T \lesssim 1$~[K], the specific heat and the thermal conductivity follow $C\propto T$ and $\kappa\propto T^2$~\cite{zeller1971thermal}, which are markedly different from the crystalline behaviors. 
These observations suggest that even far below the BP frequency, the vibrational modes are distinct from those of crystals, phonons. 
This is quite contrary to our intuition that in the continuum limit, the microscopic arrangement of particles is irrelevant so that glasses behave as the homogeneous elastic medium as do crystals, where the vibrational modes are phonons~\cite{tanguy_2002,leonforte2005continuum,monaco2009anomalous,Shiba_2016}. 

Several different phenomenological scenarios have been proposed to explain these thermal properties of glasses.
A noticeable example is the two-level systems model and the soft potential model~\cite{anderson1972anomalous, karpov1983theory,Buchenau1991,Buchenau_1992,Gurevich2003,Gurevich_2005,Gurevich_2007}. 
These theories assume that there exist the soft localized modes in glasses and that they are associated to double-well structure of the potential energy landscape.
Note that the soft localized modes have been observed in many different glassy systems~\cite{mazzacurati1996lowfrequency,Taraskin_1999,Schober_2004,Xu_2010,Baity-Jesi2015,Beltukov_2016,lerner2016statistics,Gartner_2016}.
Then, the theories consider the tunneling state in the double-well potential to explain the anomalous behaviors of specific heat and thermal conductivity of the low-temperature glasses. 
The recent version of the soft potential model takes into account the vibrational instability to explain the BP~\cite{Gurevich2003,Gurevich_2005,Gurevich_2007}.
Another important example is the theory of elastic heterogeneities~\cite{schirmacher_2006,schirmacher_2007,marruzzo2013heterogeneous,Mizuno2_2013,Mizuno_2014,Gelin2016,Zhang_2017}. 
In this theory, glasses are considered to be the elastic medium, whose elastic constants are, however, heterogeneously fluctuating in space.
These spatial fluctuations of elastic constants have been confirmed by numerical and experimental works~\cite{yoshimoto_2004,Tsamados_2009,Wagner_2011,Mizuno_2013}.
The theory uses the effective medium technique or the coherent potential approximation to treat the fluctuations of the elastic constants analytically, and then it predicts the BP as well as the anomalous sound-transport properties.

Recently, microscopic studies on the vibrational properties are developing.
Some of them focus on the simplest model of glasses, that is composed of particles interacting through the short-range, purely repulsive, harmonic potential~\cite{ohern2003jamming,wyart2005geometric,wyart2005effects,silbert2005vibrations,silbert2009normal,Wyart_2010,degiuli2014effects,Charbonneau2014,Charbonneau2014b,franz2015universal,Biroli2016,charbonneau2016universal,mizuno2017continuum}. 
An important aspect of this approach is that the mean-field theories, which become exact in infinite dimensions, are now available. 
These theories predict the low-temperature glass phase, which is characterized by the fractal free energy landscape~\cite{Charbonneau2014,Charbonneau2014b,franz2015universal,Biroli2016} or the marginal stability~\cite{wyart2005effects,Wyart_2010,degiuli2014effects}. 
Notably, the non-phonon, soft vibrational modes emerge in this glass phase, and they obey the non-Debye scaling law $g(\omega) = c \omega^2$~\cite{degiuli2014effects,franz2015universal}.
Because the prefactor $c$ is much larger than the value predicted by the Debye theory, this gives a microscopic explanation of the BP. 
These theoretical predictions were tested thoroughly by numerical simulations~\cite{charbonneau2016universal,mizuno2017continuum}. 
The non-Debye scaling $g(\omega) = c \omega^2$ indeed works well at relatively high frequency~\cite{charbonneau2016universal}. 
However, the mean-field predictions are broken down in the low-frequency, continuum limit, where the vibrational modes converge to mixture of phonon modes and soft localized modes~\cite{mizuno2017continuum}.
The phonon modes follow the Debye law, whereas the soft localized modes follow another, non-Debye scaling law $g(\omega)\propto \omega^4$.
This non-Debye scaling law is the same as the one found for several glassy systems by different means~\cite{Baity-Jesi2015,lerner2016statistics,Gartner_2016} and is also reminiscent of the soft potential model~\cite{karpov1983theory,Buchenau1991,Buchenau_1992,Gurevich2003,Gurevich_2005,Gurevich_2007}.

An important question is then what the vibrational properties are in the continuum limit for not the simplest but more realistic glasses. 
To address this question, it is natural to focus on the Lennard-Jones (LJ) system, because the LJ potential is one of the realistic model potentials for atomic systems~\cite{hansen2006theory}.
The nature of low-frequency vibrations in the LJ glasses is still controversial among previous works~\cite{leonforte2005continuum,monaco2009anomalous,lerner2016statistics,Gartner_2016}. 
Specifically, a numerical study~\cite{monaco2009anomalous} studied the mono-disperse LJ glass (in 3 dimensions) and reported that the vDOS smoothly converges to the Debye prediction below the BP frequency. 
An earlier study~\cite{leonforte2005continuum} studied the poly-disperse LJ glass and found that the frequencies of vibrational modes are discretized in the low frequency region and their levels are perfectly consistent with values predicted by the Debye theory. 
These results indicate that the vibrational modes are only phonons in the continuum limit, which is clearly different situation from that of the harmonic potential system~\cite{mizuno2017continuum}. 
On the other side, more recent works~\cite{lerner2016statistics,Gartner_2016} studied the Kob-Andersen LJ potential system and the inverse power-law  (IPL) potential system, by suppressing the effects of phonons.
In contrast to Refs.~\cite{leonforte2005continuum,monaco2009anomalous}, they observed that the soft localized modes persist below the BP frequency, following the non-Debye scaling law $g(\omega)\propto\omega^4$. 
This observation is more consistent with results of the harmonic potential system~\cite{mizuno2017continuum,lerner2016statistics}. 

To settle down these contradictory observations, we perform large-scale molecular-dynamics (MD) simulations to directly observe the low-frequency vibrational modes in the LJ glass.
In order to obtain conclusive results on the continuum limit, we simulate very large systems consisting of up to millions of particles. 
We reveal that the vibrational modes of the LJ glass are essentially the same as those of the harmonic potential system. 
In particular, we establish that the vibrational modes converge to mixture of extended phonon modes and soft localized modes in the continuum limit. 
Furthermore, we also resolve the problem why the contradictory results, described above, have been reported by previous works~\cite{leonforte2005continuum,monaco2009anomalous,lerner2016statistics,Gartner_2016}. 

The paper is organized as follows.
In Sec.~\ref{sec:methods}, we describe in detail the numerical methods including the system description, MD simulation, and the vibrational mode analysis.
Section~\ref{sec:results} contains a comprehensive presentation of our results.
We first show that the low-frequency vibrational properties are sensitive to the numerical scheme of potential truncation.
We then establish the vibrational properties in the low-frequency, continuum limit, for the ``real" LJ glass system without any potential truncation.
In Sec.~\ref{sec:discussion}, we discuss on the physical mechanism of artificial effects induced by the potential truncation.
Finally, we give our conclusion in Sec.~\ref{sec:conclusion}.

\section{Numerical methods}\label{sec:methods}
\subsection{LJ potential}\label{sec:ljpotential}
The present work focuses on a glass system composed of mono-disperse particles, which interact via the LJ pair-wise potential:
\begin{equation}
  \phi(r) = 4\epsilon\left[\left(\frac{\sigma}{r}\right)^{12} - \left(\frac{\sigma}{r}\right)^{6}\right], \label{lj}
\end{equation}
where $r$ is the distance between two particles, $\sigma$ is the diameter of particles, and $\epsilon$ represents the energy scale.
We denote mass of these particles by $m$.
When performing MD simulation on this kind of system whose potential extends over the infinite distance, we usually truncate the potential at some cutoff distance $r_c$~\cite{allen1989computer}. 
This truncation, however, makes a discontinuity at $r=r_c$, which may cause artificial effects on physical quantities.
Indeed, it has been reported that the cutoff discontinuity induces non-negligible, artificial effects on thermodynamic properties~\cite{Smit_1992,Shi_2001,Ahmed_2010} and local properties~\cite{Mizuno_2016} in some cases.

Several methods have been proposed to avoid the cutoff discontinuity in the potential. 
The simplest one is to shift the potential, such that it vanishes at $r=r_c$~\cite{allen1989computer,Xu_2012}. 
We denote this type of the LJ potential as ``LJ0":
\begin{equation}
  V_{\mathrm{LJ0}}(r) = 
  \begin{cases}
    \phi(r) - \phi(r_c) & (r<r_c), \\
    0 & (r>r_c).
  \end{cases}
\label{lj0}
\end{equation}
In the LJ0, the potential continuously vanishes at $r=r_c$, however, the first-derivative of the potential, which corresponds to the inter-particle force, still has a discontinuity.
Therefore, more sophisticated method is to shift the first-derivative of the potential as well as the potential itself, such that both of them vanish at $r=r_c$~\cite{allen1989computer,Toxvaerd_2011}. 
We denote this potential as ``LJ1":
\begin{equation}
  V_{\mathrm{LJ1}}(r) = 
  \begin{cases}
    \phi(r) - \phi(r_c) - (r-r_c)\phi'(r_c) & (r<r_c), \\
    0 & (r>r_c).
  \end{cases}
\label{lj1}
\end{equation}
Moreover, we can also smooth arbitrarily high $n$th-order ($n\ge 2$) derivatives by adding terms, $\sim (r-r_c)^n$, in the same way as in the LJ1.

We note that the previous works~\cite{leonforte2005continuum,monaco2009anomalous} have employed the same type of potentials as the LJ0 and concluded that the vibrational modes converge to phonon modes only.
On the other hand, Refs.~\cite{lerner2016statistics,Gartner_2016} used the very smooth cutoff scheme (making smooth up to the third derivative of the potential) and recognized existence of the soft localized modes below the BP frequency. 
In the present work, we study the LJ0 and LJ1 systems, and by comparing these two systems, we will conclude that the discontinuity in the inter-particle force can make crucial effects on the low-frequency vibrational modes.

\subsection{MD Simulation}\label{sec:simulation}
We perform MD simulations in $3$-dimensional, cubic boxes, under periodic boundary conditions in all directions. 
We consider two types of the LJ potentials, the LJ0~[Eq.~(\ref{lj0})] and LJ1~[Eq.~(\ref{lj1})], by changing the cutoff distance $r_c$. 
Throughout this paper, the length, temperature, and time are measured in units of $\sigma$, $\epsilon/k_B$, and $\sigma \sqrt{m/\epsilon}$, respectively ($k_B$ is the Boltzmann constant). 
The density is fixed at $\rho = N/L^3 = 0.997$ in all cases, where $N$ is the number of particles and $L$ is the linear dimension of the simulation box.
In order to access a wide range of frequencies including the continuum limit, we simulate several different system sizes, ranging from $N=8000$ to $256000$ for the LJ0 and from $N=8000$ to $1024000$ for the LJ1.

The glass configuration of LJ particles is obtained as follows. 
We first equilibrate the system in the normal liquid phase at the temperature $T=2$, by means of the microcanonical MD simulations. 
Then, using the steepest descent method, we minimize the total potential energy of the system and obtain the $T=0$ configuration of the glassy state (inherent structure). 
This protocol is equivalent to the instantaneous quench from $T=2$ (liquid phase) to $0$ (glass phase).
Note that at $T=0$, the pressure is close to zero, $P \approx 0$.
For these MD simulations, we use the program package LAMMPS~\cite{Plimpton_1995,lammps}.

\subsection{Vibrational mode analysis}\label{sec:analysis}
We next perform vibrational mode analysis on the obtained LJ glass at $T=0$.
We calculate the dynamical matrix of the LJ glass and diagonalize it to obtain the eigenvalues $\lambda_k$ and eigenvectors $\boldsymbol{e}^k=[\boldsymbol{e}^k_1,\boldsymbol{e}^k_2,\cdots,\boldsymbol{e}^k_N]$~\cite{kittel1996introduction}. 
Here, $k=1,2,\cdots,3N-3$ (three, zero-frequency translational modes are removed), and they are sorted in ascending order of eigenvalues $\lambda^k$.
The eigenvectors are normalized as $\boldsymbol{e}^k\cdot\boldsymbol{e}^k=\sum_i\boldsymbol{e}_i^k\cdot\boldsymbol{e}_i^k=1$.
From the eigenvalues, we can obtain the eigenfrequencies as $\omega_k=\sqrt{\lambda_k}$.

For the system of $N=8000$, we perform full diagonalization of the dynamical matrix to obtain all the eigenvalues and eigenvectors, by using the program package LAPACK~\cite{lapack}.
For the larger systems of $N > 8000$, we obtain the lowest eigenvalues and the associated eigenvectors, which is achieved by the program package ARPACK~\cite{arpack}.
These data from different system sizes provide the vibrational modes in the different frequency regimes, and they smoothly connect with each other.
We therefore collect these data altogether and acquire vibrational information in a wide range of frequencies including the continuum limit.

From the dataset of $\omega_k$ and $\boldsymbol{e}^k$, we calculate the following quantities to characterize the vibrational properties of the glassy system.

\subsubsection{VDOS}\label{sec:dos}
From the eigenfrequencies $\omega_k$, the vDOS is calculated as
\begin{equation}
  g(\omega) = \frac{1}{3N-3}\sum_{k=1}^{3N-3}\delta(\omega-\omega_k),
\end{equation}
where $\delta(x)$ is Dirac's delta function.
The Debye law predicts the vDOS as $g_D(\omega)= A_0 \omega^2$ with the Debye level $A_0$:
\begin{equation}
A_0 = \frac{3}{\omega_D^3},\qquad \omega_D = \left(\frac{18\pi\rho}{c_L^{-3} + 2c_T^{-3}}\right)^{1/3},
\end{equation}
where $\omega_D$ is the Debye frequency, and $c_T$ and $c_L$ are the speeds of transverse and longitudinal sound waves, respectively.
The sound speeds can be calculated from the bulk modulus $K$ and the shear modulus $G$ by the continuum mechanics, as $c_T=\sqrt{G/\rho}$ and $c_L=\sqrt{(K+2G)/\rho}$.
Here we calculated these elastic moduli by the harmonic formulation~\cite{mizuno2016elastic}.
To precisely compare our simulation data with the Debye prediction, we display our results in the form of the vDOS divided by $\omega^2$, the so called reduced vDOS $g(\omega)/\omega^2$. 

\subsubsection{Phonon order parameter}\label{sec:phononorderparameter}
To measure the extent to which each vibrational mode $k$ resembles phonons, we introduce the phonon order parameter $O^k$ as follows~\cite{mizuno2017continuum}. 
Phonons in the amorphous configuration are defined as $\boldsymbol{e}^{\boldsymbol{q},\sigma}=[\boldsymbol{e}_1^{\boldsymbol{q},\sigma},\cdots,\boldsymbol{e}_N^{\boldsymbol{q},\sigma}]$ with
\begin{equation}
  \boldsymbol{e}_i^{\boldsymbol{q},\sigma} = \frac{\boldsymbol{P}^{\boldsymbol{q},\sigma}}{\sqrt{N}}e^{j\boldsymbol{q}\cdot\boldsymbol{r}_i},
\end{equation}
where $j$ is the imaginary unit, $\boldsymbol{r}_i$ is the coordinate of $i$th particle, and $\boldsymbol{q}$ is the wavevector.
$\sigma$ represents the polarization; it takes three values, for one longitudinal ($\sigma=1$) mode and two transverse ($\sigma=2,3$) modes.
$\boldsymbol{P}^{\boldsymbol{q},\sigma}$ is the polarization vector; it is determined as $\boldsymbol{P}^{\boldsymbol{q},1} = \boldsymbol{q} /|\boldsymbol{q}|$ for longitudinal mode and $\boldsymbol{P}^{\boldsymbol{q},2} \cdot \boldsymbol{q} = \boldsymbol{P}^{\boldsymbol{q},3} \cdot \boldsymbol{q} = 0$ for two transverse modes.
Due to the finite system size of linear dimension $L$ with periodic boundary conditions, the wavevector takes discrete values as $\boldsymbol{q} = (2\pi/L)(i,j,k)$ ($i,j,k=0,1,2,\cdots$ are integers).
Note that one set of values of $\boldsymbol{q}$ and $\sigma$ uniquely determine one phonon.

We can measure the overlap between a phonon $\boldsymbol{e}^{\boldsymbol{q},\sigma}$ and a eigenvector $\boldsymbol{e}^k$ as
\begin{equation}
  \left| A^k_{\boldsymbol{q},\sigma} \right|^2 = \left| \boldsymbol{e}^k\cdot\boldsymbol{e}^{\boldsymbol{q},\sigma} \right|^2.
\end{equation}
If $\boldsymbol{e}^k$ is similar, or almost parallel, to $\boldsymbol{e}^{\boldsymbol{q},\sigma}$, then $|A^k_{\boldsymbol{q},\sigma}|^2$ becomes close to one.
We then define the phonon order parameter $O^k$ by summing up all the overlaps $\left| A^k_{\boldsymbol{q},\sigma} \right|^2$ that satisfy a certain threshold of $\left| A^k_{\boldsymbol{q},\sigma} \right|^2 > O_{\mathrm{th}}/(3N-3)$:
\begin{equation}
  \label{eq:popdef}
  O^k = \sum_{\substack{\boldsymbol{q},\sigma \\ |A^k_{\boldsymbol{q},\sigma}|^2>O_{\mathrm{th}}/(3N-3)}} |A^k_{\boldsymbol{q},\sigma}|^2.
\end{equation}
$O^k = 0$ when $\boldsymbol{e}^k$ resembles no phonons, whereas $O^k = 1$ when $\boldsymbol{e}^k$ is a phonon itself.
The choice of $O_{\mathrm{th}}$ is somewhat arbitrary; here we set $O_{\mathrm{th}} = 100$, however, we confirm that our conclusion is not affected by choice of the value $O_{\mathrm{th}}$.

\subsubsection{Participation ratio}\label{sec:participationratio}
To measure the degree of vibrational localization, we calculate the participation ratio for each mode $k$:
\begin{equation}
  P^k = \frac{1}{N}\left[\sum_{i=1}^N(\boldsymbol{e}^k_i\cdot\boldsymbol{e}^k_i)^2\right]^{-1}.
\end{equation}
This order parameter has been employed by many previous works, e.g., Refs.~\cite{mazzacurati1996lowfrequency,Taraskin_1999,Schober_2004}. 
$P^k=1/N$ when the mode $k$ involves only one particle, whereas $P^k=1$ when all the particles vibrate equally.

\section{Results}\label{sec:results}
We performed the vibrational mode analysis on the LJ1 and LJ0 glasses, as described in Sec.~\ref{sec:methods}.
We calculated the vDOS $g(\omega)$ and characterized each vibrational mode $k$ in terms of the phonon order parameter $O^k$ and the participation ratio $P^k$, which are presented below.

\begin{figure*}
  \begin{center}
    \begin{tabular}{c}
      \begin{minipage}[t]{0.5\hsize}
        \begin{center}
          \input{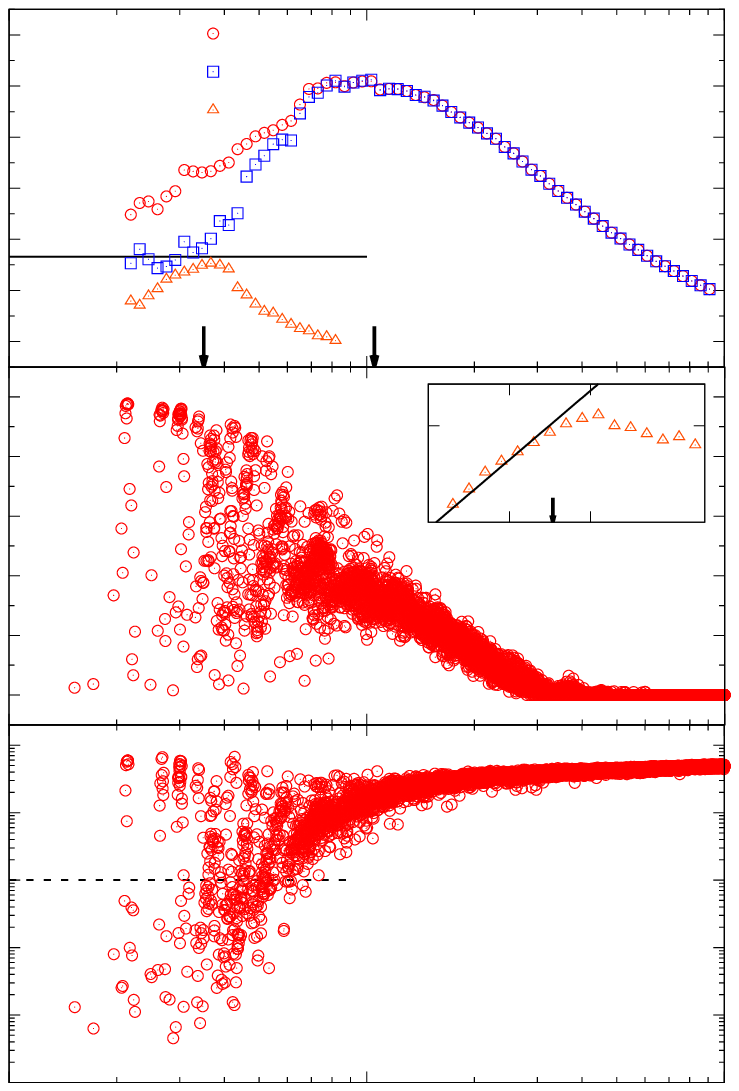}
        \end{center}
      \end{minipage}
      \begin{minipage}[t]{0.5\hsize}
        \begin{center}
          \input{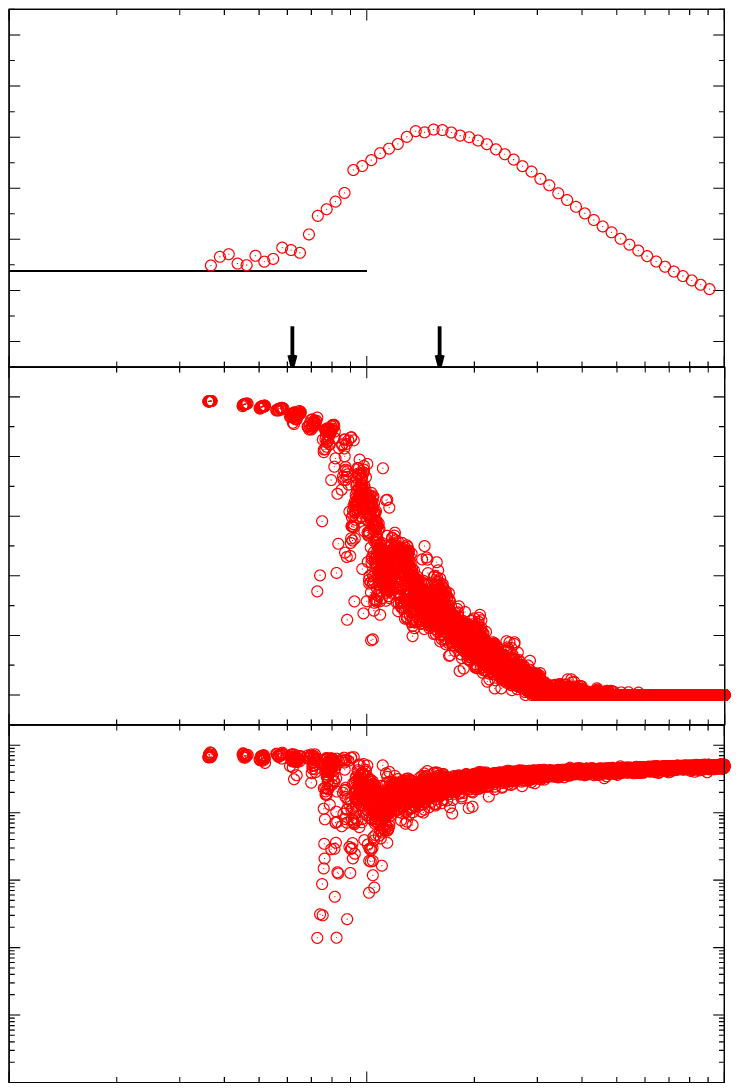}
        \end{center}
      \end{minipage}
    \end{tabular}
    \caption{
Vibrational modes in the LJ1 glass in (a) and the LJ0 glass in (b).
The panels show the reduced vDOS $g(\omega)/\omega^2$ (top), the phonon order parameter $O^k$ (middle), and the participation ratio $P^k$ (bottom).
In the top panel, we indicate the characteristic frequencies, $\omega_{\text{ex}0}$, $\omega_0$, $\omega_\text{BP}$, by arrows, and also show the Debye leve $A_0$ by the horizontal line.
The top panel in (a) shows $g(\omega)/\omega^2$ of extended modes ($P^k>10^{-2}$) and that of localized modes ($P^k<10^{-2}$).
The inset to the middle panel in (a) plots the vDOS of localized modes ($P^k<10^{-2}$).
In addition, the inset to the bottom panel in (a) shows the parametric plot of $O^k$ versus $P^k$ for different frequency regimes.
Detailed discussions on these presented data are given in the main text.
}
    \label{fig:vib}
  \end{center}
\end{figure*}

\begin{figure}
  \begin{center}
    \input{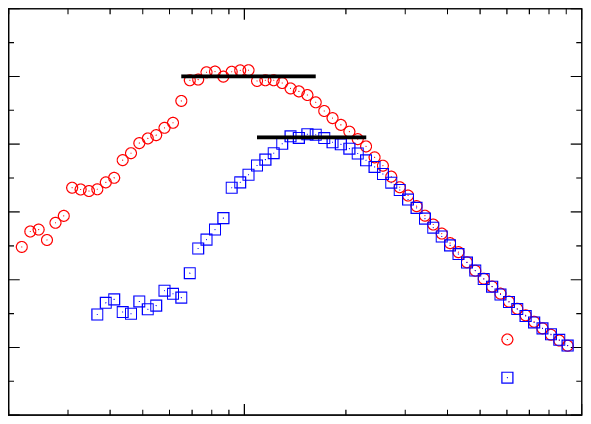}
    \caption{
Comparison of reduced vDOS between the LJ0 and LJ1 glasses.
The data are the same as those presented in Fig.~\ref{fig:vib}.
For the harmonic potential system, the plateau is predicted around $\omega_\text{BP}$ by the mean-field theories~\cite{degiuli2014effects,franz2015universal}.
Although the present work studies the LJ potential system, not the harmonic potential system, both of the LJ1 and LJ0 glasses seem to show plateaus.
The LJ1 glass shows the longer plateau than the LJ0 glass.
}
    \label{fig:boson}
  \end{center}
\end{figure}

\begin{figure}
  \begin{center}
    \input{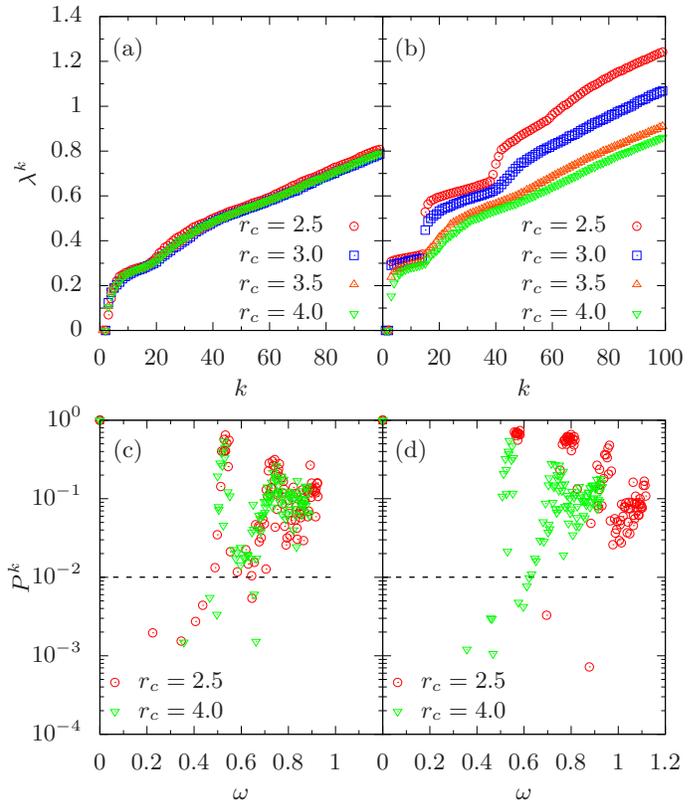}
    \caption{
Vibrational modes for several different cutoff distances $r_c$, for the LJ1 glass in (a) and (c) and the LJ0 glass in (b) and (d).
The system size is $N=64000$.
The panels (a) and (b) show the eigenvalue $\lambda^k$ as a function of the mode number $k$, for $r_c=2.5,\ 3.0,\ 3.5$, and $4.0$.
The panels (c) and (d) plot the participation ratio $P^k$ versus the eigenfrequency $\omega^k$, for $r_c=2.5$ and $4.0$.
}
    \label{fig:limit}
  \end{center}
\end{figure}

\begin{figure}
  \begin{center}
    \input{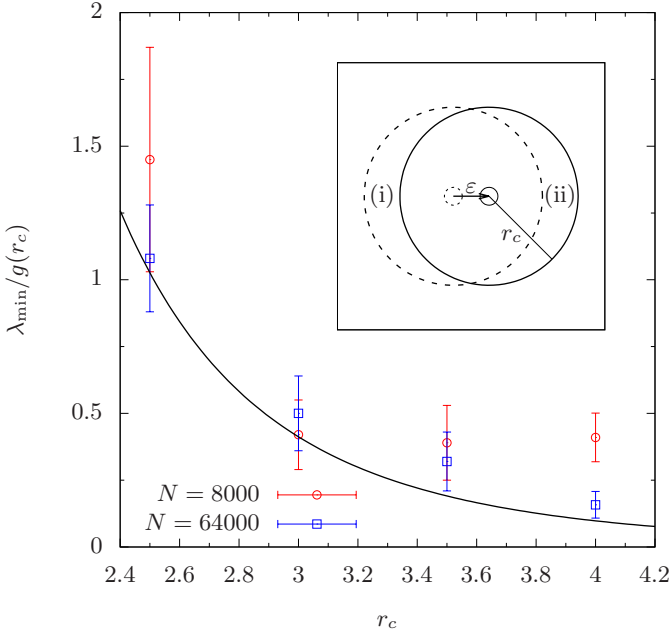}
    \caption{
Minimum eigenvalue $\lambda_{\mathrm{min}}$ in the values of $\lambda^k$ that satisfy $O^k<0.4$, for the LJ0 glass.
We plot $\lambda_{\mathrm{min}}/g(r_c)$ as a function of the cutoff length $r_c$ ($g(r_c)$ is the radial distirbution function at $r_c$).
The solid curve draws the theoretical prediction of $\lambda_{\mathrm{min}}/g(r_c) = 32\pi\rho/r_c^5$~[see Eq.~(\ref{eq:ineq})].
The inset illustrates schematic description of our argument to derive Eq.~(\ref{eq:ineq}).
}
    \label{fig:power}
  \end{center}
\end{figure}

\begin{figure}
  \begin{center}
    \input{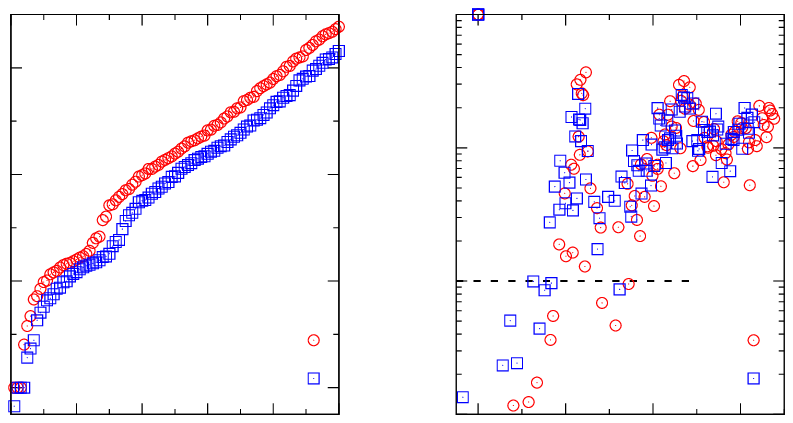}
    \caption{
Vibrational modes in the systems of 12IPL1 and 12IPL0.
The system size is $N = 64000$, and the cutoff distance is $r_c = 1.8$.
The panel (a) plots the eigenvalue $\lambda^k$ as a function of the mode number $k$.
The panel (b) shows the participation ratio $P^k$ versus the eigenvalue $\lambda^k$ for the same eigenmodes as in (a).
}
    \label{fig:softlimit}
  \end{center}
\end{figure}

\subsection{LJ1 glass}\label{sec:lj1}
We first look at the results of the LJ1 glass with the cutoff length $r_c=2.5$, which are presented in Fig.~\ref{fig:vib}(a).
As demonstrated later, these results are essentially equivalent to vibrational properties of the ``real" LJ glass, i.e., the glass of no-cutoff, infinite-range, LJ potential [Eq.~(\ref{lj})].
The top panel of Fig.~\ref{fig:vib}(a) shows the reduced vDOS $g(\omega)/\omega^2$ of the LJ1 glass.
With decreasing the frequency, $g(\omega)/\omega^2$ first increases and then starts to decrease.
As a result, it exhibits a peak, the so-called boson peak, which is located at $\omega = \omega_{\mathrm{BP}} \sim 1$.
With further decreasing $\omega$, $g(\omega)/\omega^2$ approaches the Debye level $A_0$ but does not reach it in the frequency region studied in this work.

The middle and bottom panels of Fig.~\ref{fig:vib}(a) show $O^k$ and $P^k$ for each mode $k$, respectively.
At high frequencies, $\omega \gg \omega_{\mathrm{BP}}$, the modes are characterized by non-phonon ($O^k \approx 0$) and extended ($P^k = \mathcal{O}(10^{-1})$) vibrations.
These vibrations seem to share the disordered and extended nature of floppy modes in the harmonic potential system~\cite{wyart2005geometric,wyart2005effects,silbert2005vibrations,silbert2009normal,mizuno2017continuum, Xu_2007}.
With lowering the frequency, the phonon order parameter increases up to $O^k \approx 0.3$, indicating that the modes become more phonon-like.
Remarkably around $\omega_\text{BP}$, $O^k$ bifurcates into the phonon-like group (large $O^k$) and the non-phonon group (small $O^k$).
Concomitantly, the participation ratio also bifurcates into the extended group (large $P^k$) and the localized group (small $P^k$). 
We emphasize that the phonon-like modes are extended, while the non-phonon modes are localized; the inset to the bottom panel indeed demonstrates that the modes with larger $P^k$ show larger $O^k$ and vice versa, in the lowest-frequency region of $\omega \lesssim \omega_{\text{ex}0}$ (definition of $\omega_{\text{ex}0}$ will be given later soon). 

Based on the above observation, we define the modes as extended when $P^k>10^{-2}$ and localized when $P^k<10^{-2}$.
We then calculate the vDOSs of these two types of modes, separately. 
The top panel in Fig.~\ref{fig:vib}(a) well demonstrates that $g(\omega)/\omega^2$ of the extended modes converges to the Debye level at the frequency $\omega_{\text{ex}0} \approx 0.35$ (we define $\omega_{\text{ex}0}$ here).
This means that the excess values in $g(\omega)/\omega^2$ over the Debye level at $\omega < \omega_{\text{ex}0}$ originates from the presence of the localized modes.
In addition, following the recent reports on the vDOS of localized modes~\cite{mizuno2017continuum,lerner2016statistics,Gartner_2016}, we fit the vDOS of localized modes, $g_\text{loc}(\omega)$, by the $\omega^4$ dependence.
The inset to the middle panel indeed confirms that the results are consistent with this $\omega^4$ scaling law.
We also checked that these results are not sensitive to choice of the threshold value of $P^k=10^{-2}$ to define extended and localized modes.
Therefore, all these results of the LJ1 glass are essentially the same as those of the harmonic potential system~\cite{mizuno2017continuum}, and consistent with the previous works~\cite{lerner2016statistics,Gartner_2016}. 

\subsection{LJ0 glass}\label{sec:lj0}
We next look at the results of the LJ0 glass with the cutoff length $r_c=2.5$, as presented in Fig.\ref{fig:vib}(b), where we encounter surprisingly different behaviors from the LJ1 glass.
As in the case of the LJ1 glass, with decreasing the frequency, the reduced vDOS of the LJ0 glass first increases, reaches the boson peak, and then decreases. 
However, in contrast to the LJ1 glass, $g(\omega)/\omega^2$ of the LJ0 glass does reach the Debye level $A_0$ at some finite frequency, $\omega = \omega_0 \approx 0.62$.
$O^k$ and $P^k$ more clearly show distinct behaviors of the LJ0 glass, from the LJ1 glass.
As in the LJ1 glass, $O^k$ and $P^k$ of the LJ0 glass exhibit some signatures of the bifurcation below the BP frequency $\omega_\text{BP}$. 
However remarkably, no group of the localized modes appear at $\omega \lesssim \omega_0$, and all the vibrational modes are the extended, phonon-like modes.
This is consistent with the result that $g(\omega)/\omega^2$ of the LJ0 glass exactly converges to the Debye level. 
Here, we stress that these results are consistent with the previous reports of Refs.~\cite{leonforte2005continuum,monaco2009anomalous} which employed the same cutoff treatment as in the LJ0 potential and observed the convergence to the Debye level and the phonon frequency levels at a finite frequency. 

Figure~\ref{fig:boson} compares the reduced vDOS between the LJ1 and LJ0 glasses on the equal footing.
At high frequencies, $\omega \gtrsim 3$, these two glasses exhibit almost the same values of $g(\omega)/\omega^2$.
However, we clearly see that the height of the boson peak of the LJ1 glass is higher than that of the LJ0 glass.
This difference in the height can be explained by the difference in the Debye level $A_0$; the boson peak amplitudes ($g(\omega_{\mathrm{BP}})/A_o\omega_{\mathrm{BP}}^2$) of both cases are almost the same~\cite{shintani_2008}.
More remarkably, the plateau is observed around the BP frequency $\omega_\text{BP}$ in both of the LJ glasses, as in the harmonic potential system.
For the harmonic potential system, the plateau is predicted by the mean-field theories~\cite{degiuli2014effects,franz2015universal} and confirmed by the numerical simulations~\cite{charbonneau2016universal,mizuno2017continuum}.
Although the present work studies the LJ systems, not the harmonic potential system, the LJ glasses seem to show the plateau.
This plateau is wider in the LJ1 glass than in the LJ0 glass, and  the width of the plateau in the LJ1 glass is comparable to that in the harmonic potential system.

\subsection{Infinite limit of the cutoff distance {$\boldsymbol r_c\to\infty$}}~\label{sec:thelimit}
The crucial differences between the LJ0 and LJ1 glasses raise a question: What is the nature of the low-frequency vibrational modes in the ``real" LJ glass with no-cutoff, infinite-range, LJ potential, and which type of potential is appropriate to simulate the real LJ glass?
To address this question, we discuss the infinite limit of the cutoff distance $r_c \to \infty$, because both of the LJ0 and LJ1 glasses converge to the real LJ glass in this limit.
In particular, we increase the cutoff length as $r_c=2.5,\ 3.0,\ 3.5,\ 4.0$ and performed the same vibrational mode analysis. 
For this calculation, we prepared five independent configurations of $N=64000$ for each value of $r_c$.
The results are summarized in Fig.~\ref{fig:limit}, for the LJ1 in (a) and (c) and the LJ0 in (b) and (d).
The panels (a) and (b) plot the lowest $100$ eigenvalues $\lambda^k$ as a function of the mode number $k$.
Here we averaged the eigenvalues $\lambda^k$ (at the same $k$) over the five configurations.
If the low-frequency modes are phonons (as in the case of crystals), the data should show a step-like behavior, since the phonon levels are discrete.
The LJ0 glass with $r_c=2.5$ (red circles in the panel (b)) indeed exhibits the step-like behavior up to the first two levels of phonons.
This is consistent with the results of Fig.~\ref{fig:vib}(b), where the vibratonal modes converge to the phonon modes in this system. 
However, as we make the cutoff length $r_c$ longer, the results change drastically: The step-like behavior disappears, and the data become smoother and smoother.
This clearly demonstrates that the low-frequency modes of the LJ0 glass with $r_c=2.5$ differ from those of the real LJ glass, the system with $r_c \to \infty$.

In contrast, the LJ1 glass with $r_c=2.5$ (red circles in the panel (a)) exhibits the continuous behavior. 
Importantly in the LJ1 glass, the results are rather insensitive to, or almost independent of, the value of $r_c$. 
This means that the LJ1 glass with $r_c = 2.5$ already reproduce the behavior of the real LJ glass. 

In addition, the panels (c) and (d) of Fig.~\ref{fig:limit} plot the participation ratio of the vibrational modes of the LJ0 and LJ1 glasses, for $r_c = 2.5$ and $4.0$.
We clearly see that in the LJ0 glass (the panel (d)), the soft localized modes do not appear at $r_c = 2.5$, but do appear at $r_c=4.0$.
On the other hand, the LJ1 glass (the panel (c)) shows the soft localized modes for both of $r_c = 2.5$ and $4.0$.
From these observations, we conclude that the disappearance of the soft localized modes in the LJ0 glass is due to the artificial effects caused by the discontinuity in the inter-particle force (the first derivative of the potential).
Therefore, the real LJ glass, of non-cutoff potential, should have the soft localized modes in the low-frequency (continuum) limit, $\omega \lesssim \omega_\text{ex0}$.
Our results also suggest that one has to use the LJ1 potential or the smoother potentials at the cutoff $r_c$ to study the low-frequency vibrational properties of the LJ glasses properly. 
The vibrational properties of the LJ glasses reported in the previous works~\cite{leonforte2005continuum,monaco2009anomalous} are incorrect in the low-frequency region, where the results suffer from the artificial effects of cutoff discontinuity.

\section{Discussion}\label{sec:discussion}
So far, we established that the disappearance of the soft localized modes in the LJ0 glass is not the real property of the LJ glasses, but rather this is due to the artificial effects by the improper cutoff treatment.
Then, what is the physical mechanism of this phenomenon?
To address this question, we discuss the stability of the system against the motions of particles along the soft localized modes. 
In the soft localized modes, only a few of particles carry large amplitudes of the vibrational motions. 
Here, we simply neglect the collectivity of the particle motions and regard the localized modes as motions of a tagged particle in the continuum density field of $\rho g(r)$.
Here $g(r)$ is the radial distribution function, and $r$ is the distance from the tagged particle.
This situation is schematically illustrated in the inset to Fig.~\ref{fig:power}. 
We then discuss the stability of motions of the tagged particle.

In the LJ0 glass, when the tagged particle is displaced by a infinitesimal distance $\varepsilon$, the particles in front of the tagged particle at the distance $r=r_c$ (particles in the region (ii) in the inset to Fig. \ref{fig:power}) start to exert forces on the tagged particle abruptly.
Similarly, the particles behind the tagged particles (particles in the region (i)) stop to exert forces.  
We remark that these sudden appearance and disappearance of the forces are from the artificial effects by the cut-off discontinuity of the force in the LJ0 glass, and they never emerge in the models with the smoothed force such as the LJ1 glass. 
We can estimate the total of these forces, denoted by $F_1$, in leading order with respect to $\varepsilon$: 
\begin{equation}
  \begin{split}
    F_1 &= -\rho g(r_c)\int d\Omega r_c^2 \left[-\phi'(r_c)\right]\varepsilon\cos^2\theta + \mathcal{O}(\varepsilon^2) \\
    &\simeq \frac{4\pi}{3}\rho g(r_c) r_c^2\phi'(r_c)\varepsilon \simeq \frac{32\pi\rho g(r_c)}{r_c^5}\varepsilon,
  \end{split}
\end{equation}
where $d\Omega = \sin \theta d\theta d\phi\ (0\le \theta < \pi,\ 0 \le \phi < 2\pi)$ is solid angle.
The force $F_1$ becomes positive because $\phi'(r_c)$ is positive (the LJ potential causes an attractive force at the cutoff length $r_c$). 
Therefore, this force pushes the tagged particle further away from the initial position. 
On the other side, if this motion is along an eigenmode with an eigenvalue $\lambda$, this tagged particle also feels a restoring force $F_2 = -\lambda \varepsilon$.
Then the stability of this tagged particle motion is determined by the balance of these two forces.
This argument therefore gives a mean-field estimate of the stability criteria of a soft localized mode with an eigenvalue $\lambda$, for the LJ0 glass: 
\begin{equation}
  \label{eq:ineq}
  F_1 + F_2 < 0 \ \ \Longleftrightarrow \ \ \lambda > \frac{32\pi\rho g(r_c)}{r_c^5}.
\end{equation}

To check the validity of this estimate, we study the cutoff-length $r_c$ dependence of the lowest eigenvalue $\lambda_{\mathrm{min}}$ of the localized modes with $O^k<0.4$, for the LJ0 glass.
Notice that the non-phonon modes with small $O^k$ are localized with small $P^k$ in the lowest-frequency regime (see the inset to the bottom panel in Fig.~\ref{fig:vib}(a)).
In Fig.~\ref{fig:power}, we plot $\lambda_{\mathrm{min}}$ divided by $g(r_c)$, against the cutoff length $r_c$. 
For these numerical data, we averaged  $\lambda_{\mathrm{min}}$ of five independent configurations for each case and the error bars represent the standard errors estimated from these five independent values. 
With increasing $r_c$, $\lambda_{\mathrm{min}}$ decreases, indicating that the soft localized modes extend to the lower and lower frequency regime.
Figure~\ref{fig:power} also plots the theoretical curve of $\lambda_{\mathrm{min}}/g(r_c) = 32\pi\rho/r_c^5$ in Eq.~(\ref{eq:ineq}).
The numerical data are indeed consistent with this theoretical estimation.
Note that the data of $N=8000$ become insensitive to $r_c$ at $r_c \geq 3.0$. 
We attribute this behavior to the finite system-size effect. 
Notice that data of the larger system $N=64000$ follow the theoretical curve even at $r_c \geq 3.0$, up to $r_c = 4.0$. 

If one applies the present argument to purely repulsive potential systems, the effect of cutoff discontinuity is reversed because  $\phi'(r_c)$ becomes negative (the potential causes a repulsive force at the cutoff length $r_c$).
Specifically here, we consider the $12$ inverse power-law (12IPL) potential, $\phi(r) = \epsilon ({\sigma}/{r})^{12}$, and compare the particles interacting through the shifted 12IPL potential like the LJ0 to those interacting through the force-shifted 12IPL potential like the LJ1.
We denote the shifted 12IPL potential as 12IPL0 and the force-shifted 12IPL potential as 12IPL1.
The argument of Eq.~(\ref{eq:ineq}) predicts that the soft localized modes do not disappear in the system of 12IPL0, and that even the unstable modes with negative eigenvalues can emerge.
This prediction for the 12IPL0 system is distinct from that for the LJ0 system.

To check the validity of this prediction, we perform vibrational mode analysis on the systems of 12IPL0 and 12IPL1.
We prepare configurations of $N=64000$, and the cutoff distance is fixed at $r_c = 1.8$.
The results are presented in Fig.~\ref{fig:softlimit}, where we plot the lowest $100$ eigenvalues $\lambda^k$ as a function of the mode number $k$ in (a) and the participation ratio $P^k$ versus the eigenfrequency $\omega^k$ in (b).
Differently from the LJ0 system, the 12IPL0 system does show the soft localized modes and even the unstable modes with negative eigenvalues.
On the other hand, the 12IPL1 system shows the localized modes, but no unstable modes.
These observations are fully consistent with the argument of Eq.~(\ref{eq:ineq}). 

\section{Conclusion}\label{sec:conclusion}
In the present work, we performed the large-scale MD simulations to study the continuum limit of vibrational properties of the mono-disperse LJ glasses. 
Since the LJ potential extends over the infinite range and does not have any natural cutoff, we usually introduce the truncation of the potential at some finite length $r_c$ for numerical simulations.
However, we found that this truncation can make artificial effects on the low-frequency vibrational properties.
In particular, the force discontinuity at $r_c$ improperly eliminates the localized modes in low-frequency region in the LJ glasses. 
This is the reason why the contradictory results have been reported by previous works~\cite{leonforte2005continuum,monaco2009anomalous,lerner2016statistics,Gartner_2016}.
Refs.~\cite{leonforte2005continuum,monaco2009anomalous} indeed picked up the artificial effects of the cutoff discontinuity in the inter-particle force.
Our results therefore suggest that the inter-particle force (the first derivative of the potential) be made continuous at $r = r_c$ to correctly analyze the low-frequency vibrational properties.

Our main results are then to establish vibrational properties of the ``real" LJ glass of non-cutoff, infinite-range potential.
We demonstrated that the vibrational modes converge to mixture of phonon modes and soft localized modes in the low-frequency (continuum) limit $\omega \lesssim \omega_\text{ex0}$.
The phonon modes follow the Debye law $g(\omega) = A_0 \omega^2$, whereas the soft localized modes follow another, non-Debye scaling law $g(\omega) \propto \omega^4$.
We therefore established that the realistic model of glasses (the LJ glass) shows essentially the same vibrational properties as those of the simplest model (the harmonic potential system)~\cite{mizuno2017continuum}.
Our results offer the possible universality of the low-frequency vibrational properties of glasses and amorphous materials, which do not depend on constituent molecules. 

The soft localized modes and their non-Debye scaling law are key to understand the universal thermal properties of glasses~~\cite{phillips1981amorphous,zeller1971thermal}.
This is the idea of the two-level systems model and the soft potential model~\cite{anderson1972anomalous, karpov1983theory,Buchenau1991,Buchenau_1992,Gurevich2003,Gurevich_2005,Gurevich_2007}.
The localized modes have been recognized in many types of glassy systems~~\cite{mazzacurati1996lowfrequency,Taraskin_1999,Schober_2004,Xu_2010,Beltukov_2016,Gartner_2016,lerner2016statistics}, and even in a spin glass system~\cite{Baity-Jesi2015}.
Experimentally, their existence can be anticipated as the two-level systems from the linear-temperature dependence of specific heat~\cite{Perez-Castaneda2_2014,Perez-Castaneda_2014}.
Interestingly, the most recent work~\cite{Lerner2017} reported that the nature of soft localized modes can be changed by the preparation procedure of glasses.
The soft localized modes are also key features to understand the dynamics of supercooled liquids and the yielding of amorphous materials, that are active topics focused by many previous works, e.g., Refs.~\cite{widmer_2008,Maloney2_2006,Manning_2011,Zylberg_2017}.
On the other side, the soft localized modes and the non-Debye scaling law are beyond the reach of microscopic theories developed based on the harmonic potential system~\cite{ohern2003jamming,wyart2005geometric,wyart2005effects,silbert2005vibrations,silbert2009normal,Wyart_2010,degiuli2014effects,Charbonneau2014,Charbonneau2014b,franz2015universal,Biroli2016,charbonneau2016universal,mizuno2017continuum}.
It could be an important future direction to develop the theories to capture these soft localized modes.

\begin{acknowledgments}
We thank Stefano Mossa for useful discussions and suggestions. 
This work was supported by Japan Society for the Promotion of Science (JSPS) Grant-in-Aid for Young Scientists B 17K14369, Grant-in-Aid for Young Scientists A 17H04853, and Grant-in-Aid for Scientific Research B 16H04034.
\end{acknowledgments}

\bibliographystyle{apsrev4-1}
\bibliography{draft}

\begin{thebibliography}{69}%
\makeatletter
\providecommand \@ifxundefined [1]{%
 \@ifx{#1\undefined}
}%
\providecommand \@ifnum [1]{%
 \ifnum #1\expandafter \@firstoftwo
 \else \expandafter \@secondoftwo
 \fi
}%
\providecommand \@ifx [1]{%
 \ifx #1\expandafter \@firstoftwo
 \else \expandafter \@secondoftwo
 \fi
}%
\providecommand \natexlab [1]{#1}%
\providecommand \enquote  [1]{``#1''}%
\providecommand \bibnamefont  [1]{#1}%
\providecommand \bibfnamefont [1]{#1}%
\providecommand \citenamefont [1]{#1}%
\providecommand \href@noop [0]{\@secondoftwo}%
\providecommand \href [0]{\begingroup \@sanitize@url \@href}%
\providecommand \@href[1]{\@@startlink{#1}\@@href}%
\providecommand \@@href[1]{\endgroup#1\@@endlink}%
\providecommand \@sanitize@url [0]{\catcode `\\12\catcode `\$12\catcode
  `\&12\catcode `\#12\catcode `\^12\catcode `\_12\catcode `\%12\relax}%
\providecommand \@@startlink[1]{}%
\providecommand \@@endlink[0]{}%
\providecommand \url  [0]{\begingroup\@sanitize@url \@url }%
\providecommand \@url [1]{\endgroup\@href {#1}{\urlprefix }}%
\providecommand \urlprefix  [0]{URL }%
\providecommand \Eprint [0]{\href }%
\providecommand \doibase [0]{http://dx.doi.org/}%
\providecommand \selectlanguage [0]{\@gobble}%
\providecommand \bibinfo  [0]{\@secondoftwo}%
\providecommand \bibfield  [0]{\@secondoftwo}%
\providecommand \translation [1]{[#1]}%
\providecommand \BibitemOpen [0]{}%
\providecommand \bibitemStop [0]{}%
\providecommand \bibitemNoStop [0]{.\EOS\space}%
\providecommand \EOS [0]{\spacefactor3000\relax}%
\providecommand \BibitemShut  [1]{\csname bibitem#1\endcsname}%
\let\auto@bib@innerbib\@empty
\bibitem [{\citenamefont {Phillips}(1981)}]{phillips1981amorphous}%
  \BibitemOpen
  \bibfield  {author} {\bibinfo {author} {\bibfnamefont {W.}~\bibnamefont
  {Phillips}},\ }\href {https://books.google.co.jp/books?id=EU3xAAAAMAAJ}
  {\emph {\bibinfo {title} {Amorphous solids: low-temperature properties}}},\
  Topics in current physics\ (\bibinfo  {publisher} {Springer-Verlag},\
  \bibinfo {year} {1981})\BibitemShut {NoStop}%
\bibitem [{\citenamefont {Kittel}(1996)}]{kittel1996introduction}%
  \BibitemOpen
  \bibfield  {author} {\bibinfo {author} {\bibfnamefont {C.}~\bibnamefont
  {Kittel}},\ }\href {https://books.google.co.jp/books?id=1X8pAQAAMAAJ} {\emph
  {\bibinfo {title} {Introduction to Solid State Physics}}},\ \bibinfo
  {edition} {7th}\ ed.\ (\bibinfo  {publisher} {John Wiley and Sons, New
  York},\ \bibinfo {year} {1996})\BibitemShut {NoStop}%
\bibitem [{\citenamefont {Zeller}\ and\ \citenamefont
  {Pohl}(1971)}]{zeller1971thermal}%
  \BibitemOpen
  \bibfield  {author} {\bibinfo {author} {\bibfnamefont {R.~C.}\ \bibnamefont
  {Zeller}}\ and\ \bibinfo {author} {\bibfnamefont {R.~O.}\ \bibnamefont
  {Pohl}},\ }\href {\doibase 10.1103/PhysRevB.4.2029} {\bibfield  {journal}
  {\bibinfo  {journal} {Phys. Rev. B}\ }\textbf {\bibinfo {volume} {4}},\
  \bibinfo {pages} {2029} (\bibinfo {year} {1971})}\BibitemShut {NoStop}%
\bibitem [{\citenamefont {Buchenau}\ \emph {et~al.}(1984)\citenamefont
  {Buchenau}, \citenamefont {N\"ucker},\ and\ \citenamefont
  {Dianoux}}]{buchenau1984neutron}%
  \BibitemOpen
  \bibfield  {author} {\bibinfo {author} {\bibfnamefont {U.}~\bibnamefont
  {Buchenau}}, \bibinfo {author} {\bibfnamefont {N.}~\bibnamefont {N\"ucker}},
  \ and\ \bibinfo {author} {\bibfnamefont {A.~J.}\ \bibnamefont {Dianoux}},\
  }\href {\doibase 10.1103/PhysRevLett.53.2316} {\bibfield  {journal} {\bibinfo
   {journal} {Phys. Rev. Lett.}\ }\textbf {\bibinfo {volume} {53}},\ \bibinfo
  {pages} {2316} (\bibinfo {year} {1984})}\BibitemShut {NoStop}%
\bibitem [{\citenamefont {Tanguy}\ \emph {et~al.}(2002)\citenamefont {Tanguy},
  \citenamefont {Wittmer}, \citenamefont {Leonforte},\ and\ \citenamefont
  {Barrat}}]{tanguy_2002}%
  \BibitemOpen
  \bibfield  {author} {\bibinfo {author} {\bibfnamefont {A.}~\bibnamefont
  {Tanguy}}, \bibinfo {author} {\bibfnamefont {J.~P.}\ \bibnamefont {Wittmer}},
  \bibinfo {author} {\bibfnamefont {F.}~\bibnamefont {Leonforte}}, \ and\
  \bibinfo {author} {\bibfnamefont {J.-L.}\ \bibnamefont {Barrat}},\
  }\href@noop {} {\bibfield  {journal} {\bibinfo  {journal} {Phys. Rev. B}\
  }\textbf {\bibinfo {volume} {66}},\ \bibinfo {pages} {174205} (\bibinfo
  {year} {2002})}\BibitemShut {NoStop}%
\bibitem [{\citenamefont {Leonforte}\ \emph {et~al.}(2005)\citenamefont
  {Leonforte}, \citenamefont {Boissi\`ere}, \citenamefont {Tanguy},
  \citenamefont {Wittmer},\ and\ \citenamefont
  {Barrat}}]{leonforte2005continuum}%
  \BibitemOpen
  \bibfield  {author} {\bibinfo {author} {\bibfnamefont {F.}~\bibnamefont
  {Leonforte}}, \bibinfo {author} {\bibfnamefont {R.}~\bibnamefont
  {Boissi\`ere}}, \bibinfo {author} {\bibfnamefont {A.}~\bibnamefont {Tanguy}},
  \bibinfo {author} {\bibfnamefont {J.~P.}\ \bibnamefont {Wittmer}}, \ and\
  \bibinfo {author} {\bibfnamefont {J.-L.}\ \bibnamefont {Barrat}},\ }\href
  {\doibase 10.1103/PhysRevB.72.224206} {\bibfield  {journal} {\bibinfo
  {journal} {Phys. Rev. B}\ }\textbf {\bibinfo {volume} {72}},\ \bibinfo
  {pages} {224206} (\bibinfo {year} {2005})}\BibitemShut {NoStop}%
\bibitem [{\citenamefont {Monaco}\ and\ \citenamefont
  {Mossa}(2009)}]{monaco2009anomalous}%
  \BibitemOpen
  \bibfield  {author} {\bibinfo {author} {\bibfnamefont {G.}~\bibnamefont
  {Monaco}}\ and\ \bibinfo {author} {\bibfnamefont {S.}~\bibnamefont {Mossa}},\
  }\href {\doibase 10.1073/pnas.0903922106} {\bibfield  {journal} {\bibinfo
  {journal} {Proceedings of the National Academy of Sciences}\ }\textbf
  {\bibinfo {volume} {106}},\ \bibinfo {pages} {16907} (\bibinfo {year}
  {2009})},\ \Eprint
  {http://arxiv.org/abs/http://www.pnas.org/content/106/40/16907.full.pdf}
  {http://www.pnas.org/content/106/40/16907.full.pdf} \BibitemShut {NoStop}%
\bibitem [{\citenamefont {Shiba}\ \emph {et~al.}(2016)\citenamefont {Shiba},
  \citenamefont {Yamada}, \citenamefont {Kawasaki},\ and\ \citenamefont
  {Kim}}]{Shiba_2016}%
  \BibitemOpen
  \bibfield  {author} {\bibinfo {author} {\bibfnamefont {H.}~\bibnamefont
  {Shiba}}, \bibinfo {author} {\bibfnamefont {Y.}~\bibnamefont {Yamada}},
  \bibinfo {author} {\bibfnamefont {T.}~\bibnamefont {Kawasaki}}, \ and\
  \bibinfo {author} {\bibfnamefont {K.}~\bibnamefont {Kim}},\ }\href {\doibase
  10.1103/PhysRevLett.117.245701} {\bibfield  {journal} {\bibinfo  {journal}
  {Phys. Rev. Lett.}\ }\textbf {\bibinfo {volume} {117}},\ \bibinfo {pages}
  {245701} (\bibinfo {year} {2016})}\BibitemShut {NoStop}%
\bibitem [{\citenamefont {Anderson}\ \emph {et~al.}(1972)\citenamefont
  {Anderson}, \citenamefont {Halperin},\ and\ \citenamefont
  {Varma}}]{anderson1972anomalous}%
  \BibitemOpen
  \bibfield  {author} {\bibinfo {author} {\bibfnamefont {P.~W.}\ \bibnamefont
  {Anderson}}, \bibinfo {author} {\bibfnamefont {B.~I.}\ \bibnamefont
  {Halperin}}, \ and\ \bibinfo {author} {\bibfnamefont {C.~M.}\ \bibnamefont
  {Varma}},\ }\href {\doibase 10.1080/14786437208229210} {\bibfield  {journal}
  {\bibinfo  {journal} {Philosophical Magazine}\ }\textbf {\bibinfo {volume}
  {25}},\ \bibinfo {pages} {1} (\bibinfo {year} {1972})},\ \Eprint
  {http://arxiv.org/abs/http://dx.doi.org/10.1080/14786437208229210}
  {http://dx.doi.org/10.1080/14786437208229210} \BibitemShut {NoStop}%
\bibitem [{\citenamefont {Karpov}\ \emph {et~al.}(1983)\citenamefont {Karpov},
  \citenamefont {Klinger},\ and\ \citenamefont
  {Ignat’Ev}}]{karpov1983theory}%
  \BibitemOpen
  \bibfield  {author} {\bibinfo {author} {\bibfnamefont {V.}~\bibnamefont
  {Karpov}}, \bibinfo {author} {\bibfnamefont {I.}~\bibnamefont {Klinger}}, \
  and\ \bibinfo {author} {\bibfnamefont {F.}~\bibnamefont {Ignat’Ev}},\
  }\href@noop {} {\bibfield  {journal} {\bibinfo  {journal} {Zh. Eksp. Teor.
  Fiz}\ }\textbf {\bibinfo {volume} {84}},\ \bibinfo {pages} {775} (\bibinfo
  {year} {1983})}\BibitemShut {NoStop}%
\bibitem [{\citenamefont {Buchenau}\ \emph {et~al.}(1991)\citenamefont
  {Buchenau}, \citenamefont {Galperin}, \citenamefont {Gurevich},\ and\
  \citenamefont {Schober}}]{Buchenau1991}%
  \BibitemOpen
  \bibfield  {author} {\bibinfo {author} {\bibfnamefont {U.}~\bibnamefont
  {Buchenau}}, \bibinfo {author} {\bibfnamefont {Y.~M.}\ \bibnamefont
  {Galperin}}, \bibinfo {author} {\bibfnamefont {V.~L.}\ \bibnamefont
  {Gurevich}}, \ and\ \bibinfo {author} {\bibfnamefont {H.~R.}\ \bibnamefont
  {Schober}},\ }\href {\doibase 10.1103/PhysRevB.43.5039} {\bibfield  {journal}
  {\bibinfo  {journal} {Phys. Rev. B}\ }\textbf {\bibinfo {volume} {43}},\
  \bibinfo {pages} {5039} (\bibinfo {year} {1991})}\BibitemShut {NoStop}%
\bibitem [{\citenamefont {Buchenau}\ \emph {et~al.}(1992)\citenamefont
  {Buchenau}, \citenamefont {Galperin}, \citenamefont {Gurevich}, \citenamefont
  {Parshin}, \citenamefont {Ramos},\ and\ \citenamefont
  {Schober}}]{Buchenau_1992}%
  \BibitemOpen
  \bibfield  {author} {\bibinfo {author} {\bibfnamefont {U.}~\bibnamefont
  {Buchenau}}, \bibinfo {author} {\bibfnamefont {Y.~M.}\ \bibnamefont
  {Galperin}}, \bibinfo {author} {\bibfnamefont {V.~L.}\ \bibnamefont
  {Gurevich}}, \bibinfo {author} {\bibfnamefont {D.~A.}\ \bibnamefont
  {Parshin}}, \bibinfo {author} {\bibfnamefont {M.~A.}\ \bibnamefont {Ramos}},
  \ and\ \bibinfo {author} {\bibfnamefont {H.~R.}\ \bibnamefont {Schober}},\
  }\href {\doibase 10.1103/PhysRevB.46.2798} {\bibfield  {journal} {\bibinfo
  {journal} {Phys. Rev. B}\ }\textbf {\bibinfo {volume} {46}},\ \bibinfo
  {pages} {2798} (\bibinfo {year} {1992})}\BibitemShut {NoStop}%
\bibitem [{\citenamefont {Gurevich}\ \emph {et~al.}(2003)\citenamefont
  {Gurevich}, \citenamefont {Parshin},\ and\ \citenamefont
  {Schober}}]{Gurevich2003}%
  \BibitemOpen
  \bibfield  {author} {\bibinfo {author} {\bibfnamefont {V.~L.}\ \bibnamefont
  {Gurevich}}, \bibinfo {author} {\bibfnamefont {D.~A.}\ \bibnamefont
  {Parshin}}, \ and\ \bibinfo {author} {\bibfnamefont {H.~R.}\ \bibnamefont
  {Schober}},\ }\href {\doibase 10.1103/PhysRevB.67.094203} {\bibfield
  {journal} {\bibinfo  {journal} {Phys. Rev. B}\ }\textbf {\bibinfo {volume}
  {67}},\ \bibinfo {pages} {094203} (\bibinfo {year} {2003})}\BibitemShut
  {NoStop}%
\bibitem [{\citenamefont {Gurevich}\ \emph {et~al.}(2005)\citenamefont
  {Gurevich}, \citenamefont {Parshin},\ and\ \citenamefont
  {Schober}}]{Gurevich_2005}%
  \BibitemOpen
  \bibfield  {author} {\bibinfo {author} {\bibfnamefont {V.~L.}\ \bibnamefont
  {Gurevich}}, \bibinfo {author} {\bibfnamefont {D.~A.}\ \bibnamefont
  {Parshin}}, \ and\ \bibinfo {author} {\bibfnamefont {H.~R.}\ \bibnamefont
  {Schober}},\ }\href {\doibase 10.1103/PhysRevB.71.014209} {\bibfield
  {journal} {\bibinfo  {journal} {Phys. Rev. B}\ }\textbf {\bibinfo {volume}
  {71}},\ \bibinfo {pages} {014209} (\bibinfo {year} {2005})}\BibitemShut
  {NoStop}%
\bibitem [{\citenamefont {Parshin}\ \emph {et~al.}(2007)\citenamefont
  {Parshin}, \citenamefont {Schober},\ and\ \citenamefont
  {Gurevich}}]{Gurevich_2007}%
  \BibitemOpen
  \bibfield  {author} {\bibinfo {author} {\bibfnamefont {D.~A.}\ \bibnamefont
  {Parshin}}, \bibinfo {author} {\bibfnamefont {H.~R.}\ \bibnamefont
  {Schober}}, \ and\ \bibinfo {author} {\bibfnamefont {V.~L.}\ \bibnamefont
  {Gurevich}},\ }\href {\doibase 10.1103/PhysRevB.76.064206} {\bibfield
  {journal} {\bibinfo  {journal} {Phys. Rev. B}\ }\textbf {\bibinfo {volume}
  {76}},\ \bibinfo {pages} {064206} (\bibinfo {year} {2007})}\BibitemShut
  {NoStop}%
\bibitem [{\citenamefont {Mazzacurati}\ \emph {et~al.}(1996)\citenamefont
  {Mazzacurati}, \citenamefont {Ruocco},\ and\ \citenamefont
  {Sampoli}}]{mazzacurati1996lowfrequency}%
  \BibitemOpen
  \bibfield  {author} {\bibinfo {author} {\bibfnamefont {V.}~\bibnamefont
  {Mazzacurati}}, \bibinfo {author} {\bibfnamefont {G.}~\bibnamefont {Ruocco}},
  \ and\ \bibinfo {author} {\bibfnamefont {M.}~\bibnamefont {Sampoli}},\ }\href
  {http://stacks.iop.org/0295-5075/34/i=9/a=681} {\bibfield  {journal}
  {\bibinfo  {journal} {EPL (Europhysics Letters)}\ }\textbf {\bibinfo {volume}
  {34}},\ \bibinfo {pages} {681} (\bibinfo {year} {1996})}\BibitemShut
  {NoStop}%
\bibitem [{\citenamefont {Taraskin}\ and\ \citenamefont
  {Elliott}(1999)}]{Taraskin_1999}%
  \BibitemOpen
  \bibfield  {author} {\bibinfo {author} {\bibfnamefont {S.~N.}\ \bibnamefont
  {Taraskin}}\ and\ \bibinfo {author} {\bibfnamefont {S.~R.}\ \bibnamefont
  {Elliott}},\ }\href {\doibase 10.1103/PhysRevB.59.8572} {\bibfield  {journal}
  {\bibinfo  {journal} {Phys. Rev. B}\ }\textbf {\bibinfo {volume} {59}},\
  \bibinfo {pages} {8572} (\bibinfo {year} {1999})}\BibitemShut {NoStop}%
\bibitem [{\citenamefont {Schober}\ and\ \citenamefont
  {Ruocco}(2004)}]{Schober_2004}%
  \BibitemOpen
  \bibfield  {author} {\bibinfo {author} {\bibfnamefont {H.~R.}\ \bibnamefont
  {Schober}}\ and\ \bibinfo {author} {\bibfnamefont {G.}~\bibnamefont
  {Ruocco}},\ }\href@noop {} {\bibfield  {journal} {\bibinfo  {journal}
  {Philos. Mag.}\ }\textbf {\bibinfo {volume} {84}},\ \bibinfo {pages} {1361}
  (\bibinfo {year} {2004})}\BibitemShut {NoStop}%
\bibitem [{\citenamefont {Xu}\ \emph {et~al.}(2010)\citenamefont {Xu},
  \citenamefont {Vitelli}, \citenamefont {Liu},\ and\ \citenamefont
  {Nagel}}]{Xu_2010}%
  \BibitemOpen
  \bibfield  {author} {\bibinfo {author} {\bibfnamefont {N.}~\bibnamefont
  {Xu}}, \bibinfo {author} {\bibfnamefont {V.}~\bibnamefont {Vitelli}},
  \bibinfo {author} {\bibfnamefont {A.~J.}\ \bibnamefont {Liu}}, \ and\
  \bibinfo {author} {\bibfnamefont {S.~R.}\ \bibnamefont {Nagel}},\ }\href
  {http://stacks.iop.org/0295-5075/90/i=5/a=56001} {\bibfield  {journal}
  {\bibinfo  {journal} {EPL (Europhysics Letters)}\ }\textbf {\bibinfo {volume}
  {90}},\ \bibinfo {pages} {56001} (\bibinfo {year} {2010})}\BibitemShut
  {NoStop}%
\bibitem [{\citenamefont {Baity-Jesi}\ \emph {et~al.}(2015)\citenamefont
  {Baity-Jesi}, \citenamefont {Mart\'{\i}n-Mayor}, \citenamefont {Parisi},\
  and\ \citenamefont {Perez-Gaviro}}]{Baity-Jesi2015}%
  \BibitemOpen
  \bibfield  {author} {\bibinfo {author} {\bibfnamefont {M.}~\bibnamefont
  {Baity-Jesi}}, \bibinfo {author} {\bibfnamefont {V.}~\bibnamefont
  {Mart\'{\i}n-Mayor}}, \bibinfo {author} {\bibfnamefont {G.}~\bibnamefont
  {Parisi}}, \ and\ \bibinfo {author} {\bibfnamefont {S.}~\bibnamefont
  {Perez-Gaviro}},\ }\href {\doibase 10.1103/PhysRevLett.115.267205} {\bibfield
   {journal} {\bibinfo  {journal} {Phys. Rev. Lett.}\ }\textbf {\bibinfo
  {volume} {115}},\ \bibinfo {pages} {267205} (\bibinfo {year}
  {2015})}\BibitemShut {NoStop}%
\bibitem [{\citenamefont {Beltukov}\ \emph {et~al.}(2016)\citenamefont
  {Beltukov}, \citenamefont {Fusco}, \citenamefont {Parshin},\ and\
  \citenamefont {Tanguy}}]{Beltukov_2016}%
  \BibitemOpen
  \bibfield  {author} {\bibinfo {author} {\bibfnamefont {Y.~M.}\ \bibnamefont
  {Beltukov}}, \bibinfo {author} {\bibfnamefont {C.}~\bibnamefont {Fusco}},
  \bibinfo {author} {\bibfnamefont {D.~A.}\ \bibnamefont {Parshin}}, \ and\
  \bibinfo {author} {\bibfnamefont {A.}~\bibnamefont {Tanguy}},\ }\href
  {\doibase 10.1103/PhysRevE.93.023006} {\bibfield  {journal} {\bibinfo
  {journal} {Phys. Rev. E}\ }\textbf {\bibinfo {volume} {93}},\ \bibinfo
  {pages} {023006} (\bibinfo {year} {2016})}\BibitemShut {NoStop}%
\bibitem [{\citenamefont {Lerner}\ \emph {et~al.}(2016)\citenamefont {Lerner},
  \citenamefont {D\"uring},\ and\ \citenamefont
  {Bouchbinder}}]{lerner2016statistics}%
  \BibitemOpen
  \bibfield  {author} {\bibinfo {author} {\bibfnamefont {E.}~\bibnamefont
  {Lerner}}, \bibinfo {author} {\bibfnamefont {G.}~\bibnamefont {D\"uring}}, \
  and\ \bibinfo {author} {\bibfnamefont {E.}~\bibnamefont {Bouchbinder}},\
  }\href {\doibase 10.1103/PhysRevLett.117.035501} {\bibfield  {journal}
  {\bibinfo  {journal} {Phys. Rev. Lett.}\ }\textbf {\bibinfo {volume} {117}},\
  \bibinfo {pages} {035501} (\bibinfo {year} {2016})}\BibitemShut {NoStop}%
\bibitem [{\citenamefont {Gartner}\ and\ \citenamefont
  {Lerner}(2016)}]{Gartner_2016}%
  \BibitemOpen
  \bibfield  {author} {\bibinfo {author} {\bibfnamefont {L.}~\bibnamefont
  {Gartner}}\ and\ \bibinfo {author} {\bibfnamefont {E.}~\bibnamefont
  {Lerner}},\ }\href@noop {} {\bibfield  {journal} {\bibinfo  {journal}
  {SciPost Phys.}\ }\textbf {\bibinfo {volume} {1}},\ \bibinfo {pages} {016}
  (\bibinfo {year} {2016})}\BibitemShut {NoStop}%
\bibitem [{\citenamefont {Schirmacher}(2006)}]{schirmacher_2006}%
  \BibitemOpen
  \bibfield  {author} {\bibinfo {author} {\bibfnamefont {W.}~\bibnamefont
  {Schirmacher}},\ }\href@noop {} {\bibfield  {journal} {\bibinfo  {journal}
  {Europhys. Lett.}\ }\textbf {\bibinfo {volume} {73}},\ \bibinfo {pages} {892}
  (\bibinfo {year} {2006})}\BibitemShut {NoStop}%
\bibitem [{\citenamefont {Schirmacher}\ \emph {et~al.}(2007)\citenamefont
  {Schirmacher}, \citenamefont {Ruocco},\ and\ \citenamefont
  {Scopigno}}]{schirmacher_2007}%
  \BibitemOpen
  \bibfield  {author} {\bibinfo {author} {\bibfnamefont {W.}~\bibnamefont
  {Schirmacher}}, \bibinfo {author} {\bibfnamefont {G.}~\bibnamefont {Ruocco}},
  \ and\ \bibinfo {author} {\bibfnamefont {T.}~\bibnamefont {Scopigno}},\
  }\href@noop {} {\bibfield  {journal} {\bibinfo  {journal} {Phys. Rev. Lett.}\
  }\textbf {\bibinfo {volume} {98}},\ \bibinfo {pages} {025501} (\bibinfo
  {year} {2007})}\BibitemShut {NoStop}%
\bibitem [{\citenamefont {Marruzzo}\ \emph {et~al.}(2013)\citenamefont
  {Marruzzo}, \citenamefont {Schirmacher}, \citenamefont {Fratalocchi},\ and\
  \citenamefont {Ruocco}}]{marruzzo2013heterogeneous}%
  \BibitemOpen
  \bibfield  {author} {\bibinfo {author} {\bibfnamefont {A.}~\bibnamefont
  {Marruzzo}}, \bibinfo {author} {\bibfnamefont {W.}~\bibnamefont
  {Schirmacher}}, \bibinfo {author} {\bibfnamefont {A.}~\bibnamefont
  {Fratalocchi}}, \ and\ \bibinfo {author} {\bibfnamefont {G.}~\bibnamefont
  {Ruocco}},\ }\href@noop {} {\bibfield  {journal} {\bibinfo  {journal}
  {Scientific reports}\ }\textbf {\bibinfo {volume} {3}} (\bibinfo {year}
  {2013})}\BibitemShut {NoStop}%
\bibitem [{\citenamefont {Mizuno}\ \emph
  {et~al.}(2013{\natexlab{a}})\citenamefont {Mizuno}, \citenamefont {Mossa},\
  and\ \citenamefont {Barrat}}]{Mizuno2_2013}%
  \BibitemOpen
  \bibfield  {author} {\bibinfo {author} {\bibfnamefont {H.}~\bibnamefont
  {Mizuno}}, \bibinfo {author} {\bibfnamefont {S.}~\bibnamefont {Mossa}}, \
  and\ \bibinfo {author} {\bibfnamefont {J.-L.}\ \bibnamefont {Barrat}},\
  }\href {http://stacks.iop.org/0295-5075/104/i=5/a=56001} {\bibfield
  {journal} {\bibinfo  {journal} {EPL (Europhysics Letters)}\ }\textbf
  {\bibinfo {volume} {104}},\ \bibinfo {pages} {56001} (\bibinfo {year}
  {2013}{\natexlab{a}})}\BibitemShut {NoStop}%
\bibitem [{\citenamefont {Mizuno}\ \emph {et~al.}(2014)\citenamefont {Mizuno},
  \citenamefont {Mossa},\ and\ \citenamefont {Barrat}}]{Mizuno_2014}%
  \BibitemOpen
  \bibfield  {author} {\bibinfo {author} {\bibfnamefont {H.}~\bibnamefont
  {Mizuno}}, \bibinfo {author} {\bibfnamefont {S.}~\bibnamefont {Mossa}}, \
  and\ \bibinfo {author} {\bibfnamefont {J.-L.}\ \bibnamefont {Barrat}},\
  }\href {\doibase 10.1073/pnas.1409490111} {\bibfield  {journal} {\bibinfo
  {journal} {Proceedings of the National Academy of Sciences}\ }\textbf
  {\bibinfo {volume} {111}},\ \bibinfo {pages} {11949} (\bibinfo {year}
  {2014})}\BibitemShut {NoStop}%
\bibitem [{\citenamefont {Gelin}\ \emph {et~al.}(2016)\citenamefont {Gelin},
  \citenamefont {Tanaka},\ and\ \citenamefont {Lemaitre}}]{Gelin2016}%
  \BibitemOpen
  \bibfield  {author} {\bibinfo {author} {\bibfnamefont {S.}~\bibnamefont
  {Gelin}}, \bibinfo {author} {\bibfnamefont {H.}~\bibnamefont {Tanaka}}, \
  and\ \bibinfo {author} {\bibfnamefont {A.}~\bibnamefont {Lemaitre}},\ }\href
  {http://dx.doi.org/10.1038/nmat4736} {\bibfield  {journal} {\bibinfo
  {journal} {Nature Mater.}\ }\textbf {\bibinfo {volume} {15}},\ \bibinfo
  {pages} {1177} (\bibinfo {year} {2016})}\BibitemShut {NoStop}%
\bibitem [{\citenamefont {Zhang}\ \emph {et~al.}(2017)\citenamefont {Zhang},
  \citenamefont {Zheng}, \citenamefont {Wang}, \citenamefont {Zhang},
  \citenamefont {Jin}, \citenamefont {Hong}, \citenamefont {Wang},\ and\
  \citenamefont {Zhang}}]{Zhang_2017}%
  \BibitemOpen
  \bibfield  {author} {\bibinfo {author} {\bibfnamefont {L.}~\bibnamefont
  {Zhang}}, \bibinfo {author} {\bibfnamefont {J.}~\bibnamefont {Zheng}},
  \bibinfo {author} {\bibfnamefont {Y.}~\bibnamefont {Wang}}, \bibinfo {author}
  {\bibfnamefont {L.}~\bibnamefont {Zhang}}, \bibinfo {author} {\bibfnamefont
  {Z.}~\bibnamefont {Jin}}, \bibinfo {author} {\bibfnamefont {L.}~\bibnamefont
  {Hong}}, \bibinfo {author} {\bibfnamefont {Y.}~\bibnamefont {Wang}}, \ and\
  \bibinfo {author} {\bibfnamefont {J.}~\bibnamefont {Zhang}},\ }\href@noop {}
  {\bibfield  {journal} {\bibinfo  {journal} {Nature Commun.}\ }\textbf
  {\bibinfo {volume} {8}},\ \bibinfo {pages} {67} (\bibinfo {year}
  {2017})}\BibitemShut {NoStop}%
\bibitem [{\citenamefont {Yoshimoto}\ \emph {et~al.}(2004)\citenamefont
  {Yoshimoto}, \citenamefont {Jain}, \citenamefont {VanWorkum}, \citenamefont
  {Nealey},\ and\ \citenamefont {dePablo}}]{yoshimoto_2004}%
  \BibitemOpen
  \bibfield  {author} {\bibinfo {author} {\bibfnamefont {K.}~\bibnamefont
  {Yoshimoto}}, \bibinfo {author} {\bibfnamefont {T.~S.}\ \bibnamefont {Jain}},
  \bibinfo {author} {\bibfnamefont {K.}~\bibnamefont {VanWorkum}}, \bibinfo
  {author} {\bibfnamefont {P.~F.}\ \bibnamefont {Nealey}}, \ and\ \bibinfo
  {author} {\bibfnamefont {J.~J.}\ \bibnamefont {dePablo}},\ }\href@noop {}
  {\bibfield  {journal} {\bibinfo  {journal} {Phys. Rev. Lett.}\ }\textbf
  {\bibinfo {volume} {93}},\ \bibinfo {pages} {175501} (\bibinfo {year}
  {2004})}\BibitemShut {NoStop}%
\bibitem [{\citenamefont {Tsamados}\ \emph {et~al.}(2009)\citenamefont
  {Tsamados}, \citenamefont {Tanguy}, \citenamefont {Goldenberg},\ and\
  \citenamefont {Barrat}}]{Tsamados_2009}%
  \BibitemOpen
  \bibfield  {author} {\bibinfo {author} {\bibfnamefont {M.}~\bibnamefont
  {Tsamados}}, \bibinfo {author} {\bibfnamefont {A.}~\bibnamefont {Tanguy}},
  \bibinfo {author} {\bibfnamefont {C.}~\bibnamefont {Goldenberg}}, \ and\
  \bibinfo {author} {\bibfnamefont {J.-L.}\ \bibnamefont {Barrat}},\ }\href
  {\doibase 10.1103/PhysRevE.80.026112} {\bibfield  {journal} {\bibinfo
  {journal} {Phys. Rev. E}\ }\textbf {\bibinfo {volume} {80}},\ \bibinfo
  {pages} {026112} (\bibinfo {year} {2009})}\BibitemShut {NoStop}%
\bibitem [{\citenamefont {Wagner}\ \emph {et~al.}(2011)\citenamefont {Wagner},
  \citenamefont {Bedorf}, \citenamefont {K$\ddot{\text{u}}$chemann},
  \citenamefont {Schwabe}, \citenamefont {Zhang}, \citenamefont {Arnold},\ and\
  \citenamefont {Samwer}}]{Wagner_2011}%
  \BibitemOpen
  \bibfield  {author} {\bibinfo {author} {\bibfnamefont {H.}~\bibnamefont
  {Wagner}}, \bibinfo {author} {\bibfnamefont {D.}~\bibnamefont {Bedorf}},
  \bibinfo {author} {\bibfnamefont {S.}~\bibnamefont
  {K$\ddot{\text{u}}$chemann}}, \bibinfo {author} {\bibfnamefont
  {M.}~\bibnamefont {Schwabe}}, \bibinfo {author} {\bibfnamefont
  {B.}~\bibnamefont {Zhang}}, \bibinfo {author} {\bibfnamefont
  {W.}~\bibnamefont {Arnold}}, \ and\ \bibinfo {author} {\bibfnamefont
  {K.}~\bibnamefont {Samwer}},\ }\href@noop {} {\bibfield  {journal} {\bibinfo
  {journal} {Nature Mater.}\ }\textbf {\bibinfo {volume} {10}},\ \bibinfo
  {pages} {439} (\bibinfo {year} {2011})}\BibitemShut {NoStop}%
\bibitem [{\citenamefont {Mizuno}\ \emph
  {et~al.}(2013{\natexlab{b}})\citenamefont {Mizuno}, \citenamefont {Mossa},\
  and\ \citenamefont {Barrat}}]{Mizuno_2013}%
  \BibitemOpen
  \bibfield  {author} {\bibinfo {author} {\bibfnamefont {H.}~\bibnamefont
  {Mizuno}}, \bibinfo {author} {\bibfnamefont {S.}~\bibnamefont {Mossa}}, \
  and\ \bibinfo {author} {\bibfnamefont {J.-L.}\ \bibnamefont {Barrat}},\
  }\href {\doibase 10.1103/PhysRevE.87.042306} {\bibfield  {journal} {\bibinfo
  {journal} {Phys. Rev. E}\ }\textbf {\bibinfo {volume} {87}},\ \bibinfo
  {pages} {042306} (\bibinfo {year} {2013}{\natexlab{b}})}\BibitemShut
  {NoStop}%
\bibitem [{\citenamefont {O'Hern}\ \emph {et~al.}(2003)\citenamefont {O'Hern},
  \citenamefont {Silbert}, \citenamefont {Liu},\ and\ \citenamefont
  {Nagel}}]{ohern2003jamming}%
  \BibitemOpen
  \bibfield  {author} {\bibinfo {author} {\bibfnamefont {C.~S.}\ \bibnamefont
  {O'Hern}}, \bibinfo {author} {\bibfnamefont {L.~E.}\ \bibnamefont {Silbert}},
  \bibinfo {author} {\bibfnamefont {A.~J.}\ \bibnamefont {Liu}}, \ and\
  \bibinfo {author} {\bibfnamefont {S.~R.}\ \bibnamefont {Nagel}},\ }\href
  {\doibase 10.1103/PhysRevE.68.011306} {\bibfield  {journal} {\bibinfo
  {journal} {Phys. Rev. E}\ }\textbf {\bibinfo {volume} {68}},\ \bibinfo
  {pages} {011306} (\bibinfo {year} {2003})}\BibitemShut {NoStop}%
\bibitem [{\citenamefont {Wyart}\ \emph
  {et~al.}(2005{\natexlab{a}})\citenamefont {Wyart}, \citenamefont {Nagel},\
  and\ \citenamefont {Witten}}]{wyart2005geometric}%
  \BibitemOpen
  \bibfield  {author} {\bibinfo {author} {\bibfnamefont {M.}~\bibnamefont
  {Wyart}}, \bibinfo {author} {\bibfnamefont {S.~R.}\ \bibnamefont {Nagel}}, \
  and\ \bibinfo {author} {\bibfnamefont {T.~A.}\ \bibnamefont {Witten}},\
  }\href {http://stacks.iop.org/0295-5075/72/i=3/a=486} {\bibfield  {journal}
  {\bibinfo  {journal} {EPL (Europhysics Letters)}\ }\textbf {\bibinfo {volume}
  {72}},\ \bibinfo {pages} {486} (\bibinfo {year}
  {2005}{\natexlab{a}})}\BibitemShut {NoStop}%
\bibitem [{\citenamefont {Wyart}\ \emph
  {et~al.}(2005{\natexlab{b}})\citenamefont {Wyart}, \citenamefont {Silbert},
  \citenamefont {Nagel},\ and\ \citenamefont {Witten}}]{wyart2005effects}%
  \BibitemOpen
  \bibfield  {author} {\bibinfo {author} {\bibfnamefont {M.}~\bibnamefont
  {Wyart}}, \bibinfo {author} {\bibfnamefont {L.~E.}\ \bibnamefont {Silbert}},
  \bibinfo {author} {\bibfnamefont {S.~R.}\ \bibnamefont {Nagel}}, \ and\
  \bibinfo {author} {\bibfnamefont {T.~A.}\ \bibnamefont {Witten}},\ }\href
  {\doibase 10.1103/PhysRevE.72.051306} {\bibfield  {journal} {\bibinfo
  {journal} {Phys. Rev. E}\ }\textbf {\bibinfo {volume} {72}},\ \bibinfo
  {pages} {051306} (\bibinfo {year} {2005}{\natexlab{b}})}\BibitemShut
  {NoStop}%
\bibitem [{\citenamefont {Silbert}\ \emph {et~al.}(2005)\citenamefont
  {Silbert}, \citenamefont {Liu},\ and\ \citenamefont
  {Nagel}}]{silbert2005vibrations}%
  \BibitemOpen
  \bibfield  {author} {\bibinfo {author} {\bibfnamefont {L.~E.}\ \bibnamefont
  {Silbert}}, \bibinfo {author} {\bibfnamefont {A.~J.}\ \bibnamefont {Liu}}, \
  and\ \bibinfo {author} {\bibfnamefont {S.~R.}\ \bibnamefont {Nagel}},\ }\href
  {\doibase 10.1103/PhysRevLett.95.098301} {\bibfield  {journal} {\bibinfo
  {journal} {Phys. Rev. Lett.}\ }\textbf {\bibinfo {volume} {95}},\ \bibinfo
  {pages} {098301} (\bibinfo {year} {2005})}\BibitemShut {NoStop}%
\bibitem [{\citenamefont {Silbert}\ \emph {et~al.}(2009)\citenamefont
  {Silbert}, \citenamefont {Liu},\ and\ \citenamefont
  {Nagel}}]{silbert2009normal}%
  \BibitemOpen
  \bibfield  {author} {\bibinfo {author} {\bibfnamefont {L.~E.}\ \bibnamefont
  {Silbert}}, \bibinfo {author} {\bibfnamefont {A.~J.}\ \bibnamefont {Liu}}, \
  and\ \bibinfo {author} {\bibfnamefont {S.~R.}\ \bibnamefont {Nagel}},\ }\href
  {\doibase 10.1103/PhysRevE.79.021308} {\bibfield  {journal} {\bibinfo
  {journal} {Phys. Rev. E}\ }\textbf {\bibinfo {volume} {79}},\ \bibinfo
  {pages} {021308} (\bibinfo {year} {2009})}\BibitemShut {NoStop}%
\bibitem [{\citenamefont {Wyart}(2010)}]{Wyart_2010}%
  \BibitemOpen
  \bibfield  {author} {\bibinfo {author} {\bibfnamefont {M.}~\bibnamefont
  {Wyart}},\ }\href {http://stacks.iop.org/0295-5075/89/i=6/a=64001} {\bibfield
   {journal} {\bibinfo  {journal} {EPL (Europhysics Letters)}\ }\textbf
  {\bibinfo {volume} {89}},\ \bibinfo {pages} {64001} (\bibinfo {year}
  {2010})}\BibitemShut {NoStop}%
\bibitem [{\citenamefont {DeGiuli}\ \emph {et~al.}(2014)\citenamefont
  {DeGiuli}, \citenamefont {Laversanne-Finot}, \citenamefont {During},
  \citenamefont {Lerner},\ and\ \citenamefont {Wyart}}]{degiuli2014effects}%
  \BibitemOpen
  \bibfield  {author} {\bibinfo {author} {\bibfnamefont {E.}~\bibnamefont
  {DeGiuli}}, \bibinfo {author} {\bibfnamefont {A.}~\bibnamefont
  {Laversanne-Finot}}, \bibinfo {author} {\bibfnamefont {G.}~\bibnamefont
  {During}}, \bibinfo {author} {\bibfnamefont {E.}~\bibnamefont {Lerner}}, \
  and\ \bibinfo {author} {\bibfnamefont {M.}~\bibnamefont {Wyart}},\ }\href
  {\doibase 10.1039/C4SM00561A} {\bibfield  {journal} {\bibinfo  {journal}
  {Soft Matter}\ }\textbf {\bibinfo {volume} {10}},\ \bibinfo {pages} {5628}
  (\bibinfo {year} {2014})}\BibitemShut {NoStop}%
\bibitem [{\citenamefont {Charbonneau}\ \emph
  {et~al.}(2014{\natexlab{a}})\citenamefont {Charbonneau}, \citenamefont
  {Kurchan}, \citenamefont {Parisi}, \citenamefont {Urbani},\ and\
  \citenamefont {Zamponi}}]{Charbonneau2014}%
  \BibitemOpen
  \bibfield  {author} {\bibinfo {author} {\bibfnamefont {P.}~\bibnamefont
  {Charbonneau}}, \bibinfo {author} {\bibfnamefont {J.}~\bibnamefont
  {Kurchan}}, \bibinfo {author} {\bibfnamefont {G.}~\bibnamefont {Parisi}},
  \bibinfo {author} {\bibfnamefont {P.}~\bibnamefont {Urbani}}, \ and\ \bibinfo
  {author} {\bibfnamefont {F.}~\bibnamefont {Zamponi}},\ }\href
  {http://dx.doi.org/10.1038/ncomms4725} {\bibfield  {journal} {\bibinfo
  {journal} {Nat Commun}\ }\textbf {\bibinfo {volume} {5}},\  (\bibinfo {year}
  {2014}{\natexlab{a}})}\BibitemShut {NoStop}%
\bibitem [{\citenamefont {Charbonneau}\ \emph
  {et~al.}(2014{\natexlab{b}})\citenamefont {Charbonneau}, \citenamefont
  {Kurchan}, \citenamefont {Parisi}, \citenamefont {Urbani},\ and\
  \citenamefont {Zamponi}}]{Charbonneau2014b}%
  \BibitemOpen
  \bibfield  {author} {\bibinfo {author} {\bibfnamefont {P.}~\bibnamefont
  {Charbonneau}}, \bibinfo {author} {\bibfnamefont {J.}~\bibnamefont
  {Kurchan}}, \bibinfo {author} {\bibfnamefont {G.}~\bibnamefont {Parisi}},
  \bibinfo {author} {\bibfnamefont {P.}~\bibnamefont {Urbani}}, \ and\ \bibinfo
  {author} {\bibfnamefont {F.}~\bibnamefont {Zamponi}},\ }\href
  {http://stacks.iop.org/1742-5468/2014/i=10/a=P10009} {\bibfield  {journal}
  {\bibinfo  {journal} {Journal of Statistical Mechanics: Theory and
  Experiment}\ }\textbf {\bibinfo {volume} {2014}},\ \bibinfo {pages} {P10009}
  (\bibinfo {year} {2014}{\natexlab{b}})}\BibitemShut {NoStop}%
\bibitem [{\citenamefont {Franz}\ \emph {et~al.}(2015)\citenamefont {Franz},
  \citenamefont {Parisi}, \citenamefont {Urbani},\ and\ \citenamefont
  {Zamponi}}]{franz2015universal}%
  \BibitemOpen
  \bibfield  {author} {\bibinfo {author} {\bibfnamefont {S.}~\bibnamefont
  {Franz}}, \bibinfo {author} {\bibfnamefont {G.}~\bibnamefont {Parisi}},
  \bibinfo {author} {\bibfnamefont {P.}~\bibnamefont {Urbani}}, \ and\ \bibinfo
  {author} {\bibfnamefont {F.}~\bibnamefont {Zamponi}},\ }\href {\doibase
  10.1073/pnas.1511134112} {\bibfield  {journal} {\bibinfo  {journal}
  {Proceedings of the National Academy of Sciences}\ }\textbf {\bibinfo
  {volume} {112}},\ \bibinfo {pages} {14539} (\bibinfo {year} {2015})},\
  \Eprint
  {http://arxiv.org/abs/http://www.pnas.org/content/112/47/14539.full.pdf}
  {http://www.pnas.org/content/112/47/14539.full.pdf} \BibitemShut {NoStop}%
\bibitem [{\citenamefont {Biroli}\ and\ \citenamefont
  {Urbani}(2016)}]{Biroli2016}%
  \BibitemOpen
  \bibfield  {author} {\bibinfo {author} {\bibfnamefont {G.}~\bibnamefont
  {Biroli}}\ and\ \bibinfo {author} {\bibfnamefont {P.}~\bibnamefont
  {Urbani}},\ }\href {http://dx.doi.org/10.1038/nphys3845} {\bibfield
  {journal} {\bibinfo  {journal} {Nat Phys}\ }\textbf {\bibinfo {volume}
  {12}},\ \bibinfo {pages} {1130} (\bibinfo {year} {2016})}\BibitemShut
  {NoStop}%
\bibitem [{\citenamefont {Charbonneau}\ \emph {et~al.}(2016)\citenamefont
  {Charbonneau}, \citenamefont {Corwin}, \citenamefont {Parisi}, \citenamefont
  {Poncet},\ and\ \citenamefont {Zamponi}}]{charbonneau2016universal}%
  \BibitemOpen
  \bibfield  {author} {\bibinfo {author} {\bibfnamefont {P.}~\bibnamefont
  {Charbonneau}}, \bibinfo {author} {\bibfnamefont {E.~I.}\ \bibnamefont
  {Corwin}}, \bibinfo {author} {\bibfnamefont {G.}~\bibnamefont {Parisi}},
  \bibinfo {author} {\bibfnamefont {A.}~\bibnamefont {Poncet}}, \ and\ \bibinfo
  {author} {\bibfnamefont {F.}~\bibnamefont {Zamponi}},\ }\href {\doibase
  10.1103/PhysRevLett.117.045503} {\bibfield  {journal} {\bibinfo  {journal}
  {Phys. Rev. Lett.}\ }\textbf {\bibinfo {volume} {117}},\ \bibinfo {pages}
  {045503} (\bibinfo {year} {2016})}\BibitemShut {NoStop}%
\bibitem [{\citenamefont {Mizuno}\ \emph {et~al.}(2017)\citenamefont {Mizuno},
  \citenamefont {Shiba},\ and\ \citenamefont {Ikeda}}]{mizuno2017continuum}%
  \BibitemOpen
  \bibfield  {author} {\bibinfo {author} {\bibfnamefont {H.}~\bibnamefont
  {Mizuno}}, \bibinfo {author} {\bibfnamefont {H.}~\bibnamefont {Shiba}}, \
  and\ \bibinfo {author} {\bibfnamefont {A.}~\bibnamefont {Ikeda}},\ }\href
  {\doibase 10.1073/pnas.1709015114} {\bibfield  {journal} {\bibinfo  {journal}
  {Proceedings of the National Academy of Sciences}\ } (\bibinfo {year}
  {2017}),\ 10.1073/pnas.1709015114}\BibitemShut {NoStop}%
\bibitem [{\citenamefont {Hansen}\ and\ \citenamefont
  {McDonald}(2006)}]{hansen2006theory}%
  \BibitemOpen
  \bibfield  {author} {\bibinfo {author} {\bibfnamefont {J.}~\bibnamefont
  {Hansen}}\ and\ \bibinfo {author} {\bibfnamefont {I.}~\bibnamefont
  {McDonald}},\ }\href {https://books.google.co.jp/books?id=Uhm87WZBnxEC}
  {\emph {\bibinfo {title} {Theory of Simple Liquids}}}\ (\bibinfo  {publisher}
  {Elsevier Science},\ \bibinfo {year} {2006})\BibitemShut {NoStop}%
\bibitem [{\citenamefont {Allen}\ and\ \citenamefont
  {Tildesley}(1989)}]{allen1989computer}%
  \BibitemOpen
  \bibfield  {author} {\bibinfo {author} {\bibfnamefont {M.}~\bibnamefont
  {Allen}}\ and\ \bibinfo {author} {\bibfnamefont {D.}~\bibnamefont
  {Tildesley}},\ }\href {https://books.google.co.jp/books?id=O32VXB9e5P4C}
  {\emph {\bibinfo {title} {Computer Simulation of Liquids}}},\ Oxford Science
  Publ\ (\bibinfo  {publisher} {Clarendon Press},\ \bibinfo {year}
  {1989})\BibitemShut {NoStop}%
\bibitem [{\citenamefont {Smit}(1992)}]{Smit_1992}%
  \BibitemOpen
  \bibfield  {author} {\bibinfo {author} {\bibfnamefont {B.}~\bibnamefont
  {Smit}},\ }\href@noop {} {\bibfield  {journal} {\bibinfo  {journal} {The
  Journal of Chemical Physics}\ }\textbf {\bibinfo {volume} {96}} (\bibinfo
  {year} {1992})}\BibitemShut {NoStop}%
\bibitem [{\citenamefont {Shi}\ and\ \citenamefont {Johnson}(2001)}]{Shi_2001}%
  \BibitemOpen
  \bibfield  {author} {\bibinfo {author} {\bibfnamefont {W.}~\bibnamefont
  {Shi}}\ and\ \bibinfo {author} {\bibfnamefont {J.}~\bibnamefont {Johnson}},\
  }\href {\doibase http://dx.doi.org/10.1016/S0378-3812(01)00534-9} {\bibfield
  {journal} {\bibinfo  {journal} {Fluid Phase Equilibria}\ }\textbf {\bibinfo
  {volume} {187-188}},\ \bibinfo {pages} {171 } (\bibinfo {year}
  {2001})}\BibitemShut {NoStop}%
\bibitem [{\citenamefont {Ahmed}\ and\ \citenamefont
  {Sadus}(2010)}]{Ahmed_2010}%
  \BibitemOpen
  \bibfield  {author} {\bibinfo {author} {\bibfnamefont {A.}~\bibnamefont
  {Ahmed}}\ and\ \bibinfo {author} {\bibfnamefont {R.~J.}\ \bibnamefont
  {Sadus}},\ }\href {\doibase http://dx.doi.org/10.1063/1.3481102} {\bibfield
  {journal} {\bibinfo  {journal} {The Journal of Chemical Physics}\ }\textbf
  {\bibinfo {volume} {133}},\ \bibinfo {eid} {124515} (\bibinfo {year}
  {2010})}\BibitemShut {NoStop}%
\bibitem [{\citenamefont {Mizuno}\ \emph
  {et~al.}(2016{\natexlab{a}})\citenamefont {Mizuno}, \citenamefont {Silbert},
  \citenamefont {Sperl}, \citenamefont {Mossa},\ and\ \citenamefont
  {Barrat}}]{Mizuno_2016}%
  \BibitemOpen
  \bibfield  {author} {\bibinfo {author} {\bibfnamefont {H.}~\bibnamefont
  {Mizuno}}, \bibinfo {author} {\bibfnamefont {L.~E.}\ \bibnamefont {Silbert}},
  \bibinfo {author} {\bibfnamefont {M.}~\bibnamefont {Sperl}}, \bibinfo
  {author} {\bibfnamefont {S.}~\bibnamefont {Mossa}}, \ and\ \bibinfo {author}
  {\bibfnamefont {J.-L.}\ \bibnamefont {Barrat}},\ }\href@noop {} {\bibfield
  {journal} {\bibinfo  {journal} {Phys. Rev. E}\ }\textbf {\bibinfo {volume}
  {93}} (\bibinfo {year} {2016}{\natexlab{a}})}\BibitemShut {NoStop}%
\bibitem [{\citenamefont {Xu}\ \emph {et~al.}(2012)\citenamefont {Xu},
  \citenamefont {Wittmer}, \citenamefont {Poli\ifmmode~\acute{n}\else
  \'{n}\fi{}ska},\ and\ \citenamefont {Baschnagel}}]{Xu_2012}%
  \BibitemOpen
  \bibfield  {author} {\bibinfo {author} {\bibfnamefont {H.}~\bibnamefont
  {Xu}}, \bibinfo {author} {\bibfnamefont {J.~P.}\ \bibnamefont {Wittmer}},
  \bibinfo {author} {\bibfnamefont {P.}~\bibnamefont
  {Poli\ifmmode~\acute{n}\else \'{n}\fi{}ska}}, \ and\ \bibinfo {author}
  {\bibfnamefont {J.}~\bibnamefont {Baschnagel}},\ }\href {\doibase
  10.1103/PhysRevE.86.046705} {\bibfield  {journal} {\bibinfo  {journal} {Phys.
  Rev. E}\ }\textbf {\bibinfo {volume} {86}},\ \bibinfo {pages} {046705}
  (\bibinfo {year} {2012})}\BibitemShut {NoStop}%
\bibitem [{\citenamefont {Toxvaerd}\ and\ \citenamefont
  {Dyre}(2011)}]{Toxvaerd_2011}%
  \BibitemOpen
  \bibfield  {author} {\bibinfo {author} {\bibfnamefont {S.}~\bibnamefont
  {Toxvaerd}}\ and\ \bibinfo {author} {\bibfnamefont {J.~C.}\ \bibnamefont
  {Dyre}},\ }\href {\doibase http://dx.doi.org/10.1063/1.3558787} {\bibfield
  {journal} {\bibinfo  {journal} {The Journal of Chemical Physics}\ }\textbf
  {\bibinfo {volume} {134}},\ \bibinfo {eid} {081102} (\bibinfo {year}
  {2011})}\BibitemShut {NoStop}%
\bibitem [{\citenamefont {Plimpton}(1995)}]{Plimpton_1995}%
  \BibitemOpen
  \bibfield  {author} {\bibinfo {author} {\bibfnamefont {S.}~\bibnamefont
  {Plimpton}},\ }\href {\doibase http://dx.doi.org/10.1006/jcph.1995.1039}
  {\bibfield  {journal} {\bibinfo  {journal} {Journal of Computational
  Physics}\ }\textbf {\bibinfo {volume} {117}},\ \bibinfo {pages} {1 }
  (\bibinfo {year} {1995})}\BibitemShut {NoStop}%
\bibitem [{lam()}]{lammps}%
  \BibitemOpen
  \href@noop {} {}\bibinfo {note} {Http://lammps.sandia.gov.}\BibitemShut
  {Stop}%
\bibitem [{lap()}]{lapack}%
  \BibitemOpen
  \href@noop {} {}\bibinfo {note} {Http://www.netlib.org/lapack/.}\BibitemShut
  {Stop}%
\bibitem [{arp()}]{arpack}%
  \BibitemOpen
  \href@noop {} {}\bibinfo {note}
  {Http://www.caam.rice.edu/software/ARPACK.}\BibitemShut {Stop}%
\bibitem [{\citenamefont {Mizuno}\ \emph
  {et~al.}(2016{\natexlab{b}})\citenamefont {Mizuno}, \citenamefont {Saitoh},\
  and\ \citenamefont {Silbert}}]{mizuno2016elastic}%
  \BibitemOpen
  \bibfield  {author} {\bibinfo {author} {\bibfnamefont {H.}~\bibnamefont
  {Mizuno}}, \bibinfo {author} {\bibfnamefont {K.}~\bibnamefont {Saitoh}}, \
  and\ \bibinfo {author} {\bibfnamefont {L.~E.}\ \bibnamefont {Silbert}},\
  }\href {\doibase 10.1103/PhysRevE.93.062905} {\bibfield  {journal} {\bibinfo
  {journal} {Phys. Rev. E}\ }\textbf {\bibinfo {volume} {93}},\ \bibinfo
  {pages} {062905} (\bibinfo {year} {2016}{\natexlab{b}})}\BibitemShut
  {NoStop}%
\bibitem [{\citenamefont {Xu}\ \emph {et~al.}(2007)\citenamefont {Xu},
  \citenamefont {Wyart}, \citenamefont {Liu},\ and\ \citenamefont
  {Nagel}}]{Xu_2007}%
  \BibitemOpen
  \bibfield  {author} {\bibinfo {author} {\bibfnamefont {N.}~\bibnamefont
  {Xu}}, \bibinfo {author} {\bibfnamefont {M.}~\bibnamefont {Wyart}}, \bibinfo
  {author} {\bibfnamefont {A.~J.}\ \bibnamefont {Liu}}, \ and\ \bibinfo
  {author} {\bibfnamefont {S.~R.}\ \bibnamefont {Nagel}},\ }\href {\doibase
  10.1103/PhysRevLett.98.175502} {\bibfield  {journal} {\bibinfo  {journal}
  {Phys. Rev. Lett.}\ }\textbf {\bibinfo {volume} {98}},\ \bibinfo {pages}
  {175502} (\bibinfo {year} {2007})}\BibitemShut {NoStop}%
\bibitem [{\citenamefont {Shintani}\ and\ \citenamefont
  {Tanaka}(2008)}]{shintani_2008}%
  \BibitemOpen
  \bibfield  {author} {\bibinfo {author} {\bibfnamefont {H.}~\bibnamefont
  {Shintani}}\ and\ \bibinfo {author} {\bibfnamefont {H.}~\bibnamefont
  {Tanaka}},\ }\href@noop {} {\bibfield  {journal} {\bibinfo  {journal} {Nature
  Mater.}\ }\textbf {\bibinfo {volume} {7}},\ \bibinfo {pages} {870} (\bibinfo
  {year} {2008})}\BibitemShut {NoStop}%
\bibitem [{\citenamefont {Pérez-Castañeda}\ \emph
  {et~al.}(2014{\natexlab{a}})\citenamefont {Pérez-Castañeda}, \citenamefont
  {Jiménez-Rioboo},\ and\ \citenamefont {Ramos}}]{Perez-Castaneda2_2014}%
  \BibitemOpen
  \bibfield  {author} {\bibinfo {author} {\bibfnamefont {T.}~\bibnamefont
  {Pérez-Castañeda}}, \bibinfo {author} {\bibfnamefont {R.~J.}\ \bibnamefont
  {Jiménez-Rioboo}}, \ and\ \bibinfo {author} {\bibfnamefont {M.~A.}\
  \bibnamefont {Ramos}},\ }\href@noop {} {\bibfield  {journal} {\bibinfo
  {journal} {Phys. Rev. Lett.}\ }\textbf {\bibinfo {volume} {112}},\ \bibinfo
  {pages} {165901} (\bibinfo {year} {2014}{\natexlab{a}})}\BibitemShut
  {NoStop}%
\bibitem [{\citenamefont {Pérez-Castañeda}\ \emph
  {et~al.}(2014{\natexlab{b}})\citenamefont {Pérez-Castañeda}, \citenamefont
  {Rodríguez-Tinoco}, \citenamefont {Rodríguez-Viejo},\ and\ \citenamefont
  {Ramos}}]{Perez-Castaneda_2014}%
  \BibitemOpen
  \bibfield  {author} {\bibinfo {author} {\bibfnamefont {T.}~\bibnamefont
  {Pérez-Castañeda}}, \bibinfo {author} {\bibfnamefont {C.}~\bibnamefont
  {Rodríguez-Tinoco}}, \bibinfo {author} {\bibfnamefont {J.}~\bibnamefont
  {Rodríguez-Viejo}}, \ and\ \bibinfo {author} {\bibfnamefont {M.~A.}\
  \bibnamefont {Ramos}},\ }\href {\doibase 10.1073/pnas.1405545111} {\bibfield
  {journal} {\bibinfo  {journal} {Proceedings of the National Academy of
  Sciences}\ }\textbf {\bibinfo {volume} {111}},\ \bibinfo {pages} {11275}
  (\bibinfo {year} {2014}{\natexlab{b}})}\BibitemShut {NoStop}%
\bibitem [{\citenamefont {Lerner}\ and\ \citenamefont
  {Bouchbinder}(2017)}]{Lerner2017}%
  \BibitemOpen
  \bibfield  {author} {\bibinfo {author} {\bibfnamefont {E.}~\bibnamefont
  {Lerner}}\ and\ \bibinfo {author} {\bibfnamefont {E.}~\bibnamefont
  {Bouchbinder}},\ }\href {\doibase 10.1103/PhysRevE.96.020104} {\bibfield
  {journal} {\bibinfo  {journal} {Phys. Rev. E}\ }\textbf {\bibinfo {volume}
  {96}},\ \bibinfo {pages} {020104} (\bibinfo {year} {2017})}\BibitemShut
  {NoStop}%
\bibitem [{\citenamefont {Widmer-Cooper}\ \emph {et~al.}(2008)\citenamefont
  {Widmer-Cooper}, \citenamefont {Perry}, \citenamefont {Harrowell},\ and\
  \citenamefont {Reichman}}]{widmer_2008}%
  \BibitemOpen
  \bibfield  {author} {\bibinfo {author} {\bibfnamefont {A.}~\bibnamefont
  {Widmer-Cooper}}, \bibinfo {author} {\bibfnamefont {H.}~\bibnamefont
  {Perry}}, \bibinfo {author} {\bibfnamefont {P.}~\bibnamefont {Harrowell}}, \
  and\ \bibinfo {author} {\bibfnamefont {D.~R.}\ \bibnamefont {Reichman}},\
  }\href@noop {} {\bibfield  {journal} {\bibinfo  {journal} {Nature phys.}\
  }\textbf {\bibinfo {volume} {4}},\ \bibinfo {pages} {711} (\bibinfo {year}
  {2008})}\BibitemShut {NoStop}%
\bibitem [{\citenamefont {Maloney}\ and\ \citenamefont
  {Lema\^{\i}tre}(2006)}]{Maloney2_2006}%
  \BibitemOpen
  \bibfield  {author} {\bibinfo {author} {\bibfnamefont {C.~E.}\ \bibnamefont
  {Maloney}}\ and\ \bibinfo {author} {\bibfnamefont {A.}~\bibnamefont
  {Lema\^{\i}tre}},\ }\href {\doibase 10.1103/PhysRevE.74.016118} {\bibfield
  {journal} {\bibinfo  {journal} {Phys. Rev. E}\ }\textbf {\bibinfo {volume}
  {74}},\ \bibinfo {pages} {016118} (\bibinfo {year} {2006})}\BibitemShut
  {NoStop}%
\bibitem [{\citenamefont {Manning}\ and\ \citenamefont
  {Liu}(2011)}]{Manning_2011}%
  \BibitemOpen
  \bibfield  {author} {\bibinfo {author} {\bibfnamefont {M.~L.}\ \bibnamefont
  {Manning}}\ and\ \bibinfo {author} {\bibfnamefont {A.~J.}\ \bibnamefont
  {Liu}},\ }\href {\doibase 10.1103/PhysRevLett.107.108302} {\bibfield
  {journal} {\bibinfo  {journal} {Phys. Rev. Lett.}\ }\textbf {\bibinfo
  {volume} {107}},\ \bibinfo {pages} {108302} (\bibinfo {year}
  {2011})}\BibitemShut {NoStop}%
\bibitem [{\citenamefont {Zylberg}\ \emph {et~al.}(2017)\citenamefont
  {Zylberg}, \citenamefont {Lerner}, \citenamefont {Bar-Sinai},\ and\
  \citenamefont {Bouchbinder}}]{Zylberg_2017}%
  \BibitemOpen
  \bibfield  {author} {\bibinfo {author} {\bibfnamefont {J.}~\bibnamefont
  {Zylberg}}, \bibinfo {author} {\bibfnamefont {E.}~\bibnamefont {Lerner}},
  \bibinfo {author} {\bibfnamefont {Y.}~\bibnamefont {Bar-Sinai}}, \ and\
  \bibinfo {author} {\bibfnamefont {E.}~\bibnamefont {Bouchbinder}},\ }\href
  {\doibase 10.1073/pnas.1704403114} {\bibfield  {journal} {\bibinfo  {journal}
  {Proceedings of the National Academy of Sciences}\ }\textbf {\bibinfo
  {volume} {114}},\ \bibinfo {pages} {7289} (\bibinfo {year} {2017})},\ \Eprint
  {http://arxiv.org/abs/http://www.pnas.org/content/114/28/7289.full.pdf}
  {http://www.pnas.org/content/114/28/7289.full.pdf} \BibitemShut {NoStop}%
\end{thebibliography}%

\end{document}